\documentclass[aps,prb,twocolumn]{revtex4}
\usepackage{color}
\usepackage{float}
\usepackage{amsmath}
\usepackage{amssymb}
\usepackage{graphicx}
\begin{document}
\title{Conductance response of graphene nanoribbons and quantum point contacts in the scanning gate measurements}
\author{A. Mre\'nca, K. Kolasi\'nski, and B. Szafran}
\affiliation{AGH University of Science and Technology, Faculty of Physics and Applied Computer Science,\\
al. Mickiewicza 30, 30-059 Krak\'ow, Poland}
\begin{abstract}
We provide a theoretical study of the conductance response of systems based on graphene nanoribbon to the potential
of a scanning probe. The study is based on the Landauer approach for the tight-binding
Hamiltonian with an implementation of the quantum transmitting boundary method and covers homogenous nanoribbons, their asymmetric narrowing  and quantum point contacts of various profiles.
The response maps at low Fermi energies resolve formation of n-p junctions induced by the probe potential
and a presence of zigzag-armchair segments of the edges for inhomogeneous ribbons. For an asymmetric narrowing
of the nanoribbons the scanning probe resolves formation of standing waves
related to backscattering within the highest subband of the narrower part of the system.
The QPCs containing a long constriction support formation of localized resonances which
induce a system of conductance peaks that is reentrant in the Fermi energy, with the form of the probability density
that can be resolved by the conductance mapping. For shorter constrictions the probe induces smooth conductance minima
within the constrictions. In general, besides the low-energy transport gap, in the wider parts of the ribbon 
the variation of the conductance of the system is low compared to the narrower part.
\end{abstract}
\maketitle

\section{Introduction}
The electron flow in semiconductor nanostructures containing a two-dimensional electron gas (2DEG) buried shallow beneath the
surface of the structure can be probed by a charged tip of an atomic force microscope. The technique, known
as the scanning gate microscopy \cite{sgmr} (SGM), gathers conductance maps as functions of the position of the tip that is capacitively
coupled to the electron gas.
For 2DEG confined in bulk semiconductors the maps are used to extract electron trajectories \cite{traj},
branching of the electron flow via quantum point contacts (QPCs) \cite{qpcsb}, many-body phenomena \cite{brun},  Aharonov-Bohm effects \cite{qr}
and scarred wave functions in quantum billiards \cite{bil} as well as the interference effects \cite{qpcsb} due to backscattering by the tip.
The SGM technique has also been used \cite{b3,c2,c3,c1,b1,b2,grring} for graphene with the electron gas remaining strictly on the surface of the structure.
In particular, SGM technique was applied in studies of charge inhomogeneities\cite{b3,c2},
charging localized states formed within a widening of the nanoribbons \cite{c3} and at potential constrictions \cite{c1}, universal conductance fluctuations \cite{b1} and weak localization effects \cite{b2}.
Very recently a SGM study of the graphene quantum rings was reported \cite{grring} with conductance fringes as due to resonant localized states.

Quantum point contacts are basic elements of the quantum transport circuitry allowing for the current injection and detection, control of the number of conducting of modes, etc. 
The conductance quantization for QPCs defined within the bulk semiconductors is usually observed for electrostatic  potentials \cite{qpcq} controlling the width of the constriction.
For graphene systems this approach is rather excluded, since 
the Klein tunneling \cite{kt,kk} and perfectly conducting channels \cite{pcc,pcc2}
in graphene nanoribbons lead to a low effectiveness of electrostatic confinement and backscattering,
allowing the Dirac electrons to pass across potential barriers \cite{abnormal}.
The QPCs are therefore formed by tailoring constrictions \cite{lin,tombros,bialorusy} of the nanoribbons \cite{nrb1,nrb2,nrb3,nrb4,nrb5,nrb6,nrb7,nrb8}.
The ineffectiveness of backscattering in graphene seems reflected by the results of the experimental SGM studies of the quantum point contacts (QPC) \cite{qpce,c1} which find
flat conductance maps outside the QPC  \cite{qpce}
 in a distinct contrast to the results obtained for bulk semiconductor in which interference fringes due to the backscattering by the tip and the resulting formation
of standing wave between the QPC and the probe are clearly observed \cite{qpcsb}.
The scanning gate microscopy was used \cite{aplpa} for detection of spontaneous quantum dots formed along the disordered ribbon within the transport gap \cite{nrb1}.

The purpose of the present paper is to determine the response of graphene quantum point contacts formed by constrictions of the nanoribbons to the scanning probe [Fig. 1(c)]. 
As a starting point of the study we consider reaction of homogenous graphene nanoribbons [Fig. 1(a)] to the perturbation by an external potential
and asymmetric narrowing [Fig. 1 (b)] of the channels \cite{wurmnjp}. The QPC [Fig. 1(c)] is formed by two asymmetric connections [Fig. 1(b)], whose scattering 
properties and response to the probe needs to be determined in order to separate the effects of the resonances formed within the QPC. 
We consider systems of perfect armchair or zigzag edges -- and systems varying with atomic-step edges,
as can be produced with the Joule heating technique \cite{jule}.  For pristine graphene ribbons \cite{nrb1,nrb2,nrb3} the edge determines the character
of the ribbon (semiconducting or semimetallic) and the ribbons with boundaries other than armchair have the transport properties similar to the zigzag edges \cite{43}.
For the purpose of the present study we develop an implementation of the quantum transmitting boundary \cite{bib1} method for the solution of the coherent transport problem set by scattering solution
of the Schr\"odinger equation for the tight-binding Hamiltonian. We study both the effects of the local potential variation induced by the tip on edge-localized states
and the effects of backscattering within the constrictions.
For QPCs the scattering induced by disorder \cite{qpcsb} and the inhomogeneity of the edges \cite{qpcd} destroy the conductance quantization which can still be observed  \cite{tombros}
for mesoscopically smooth boundaries \cite{bialorusy} of the constriction.
As compared to QPCs, we find that an asymmetric narrowing, i.e. a contact between nanoribbons of varied width, exhibits a very clear conductance quantization as a function of the Fermi energy, also
when the edges are not smooth at the mesoscopic scale.
We study quantum point contacts and find that for short constrictions there is a flat minimum of conductance
for the tip above the constriction in agreement with the results of Ref. \cite{qpce}. Generally, for any tip potential we find that outside the graphene constrictions the variation of the conductance by the tip potential is not pronounced as in the experiments of Refs. \cite{qpce,c1}.
For longer constrictions however a regular formation of the standing waves occurs which -- according to the present study -- should be resolved by the scanning gate microscopy.

The resonant states localized at armchair-zigzag connections at ribbon constrictions formed
by atomic steps \cite{tg,bialorusy} induce a strong backscattering. These resonant states quench the conductance to zero at low energy. We find that the charge transport can be unblocked by the potential of the probe
which neutralizes the scattering centers, with conductance maps forming halos similar to those found in the Coulomb blockade experiment of Ref. \cite{c1}. The response  for homogenous nanoribbons to the probe potential at low Fermi energies is determined by the existence of perfectly conducting channels (metallic armchair and zigzag ribbons), as well as by
formation of local n-p junctions at the edges (zigzag ribbons).

\begin{figure}[htbp]
\begin{center}
 \includegraphics[width=0.45\textwidth]{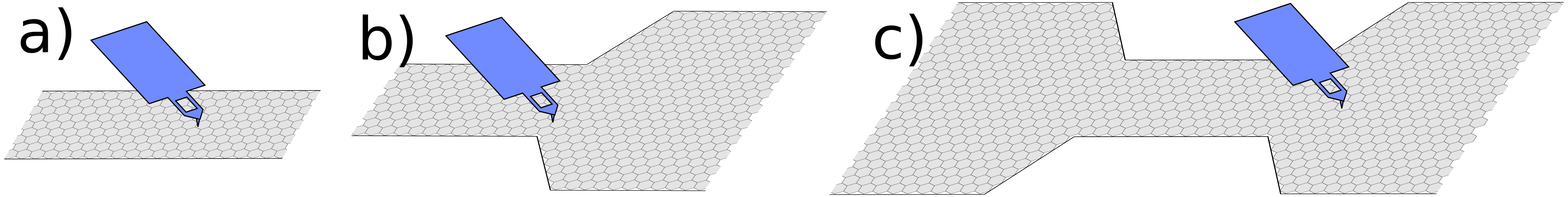}
\caption{ Schematic drawing of the systems studied: nanoribbon (a), asymmetric narrowing (b) and QPC (c).}\label{schema}
\end{center}
\end{figure}

This paper is organized as follows: in Section II we outline the calculation method which is described in detail in the Appendix, Section III, IV and V describe
the response of the nanoribbons, their narrowing and constrictions, respectively (see Fig. \ref{schema}). Summary and conclusions are given in Section VI.

\section{Method}

\begin{figure}[H]
\begin{center}
 \includegraphics[width=0.3\textwidth]{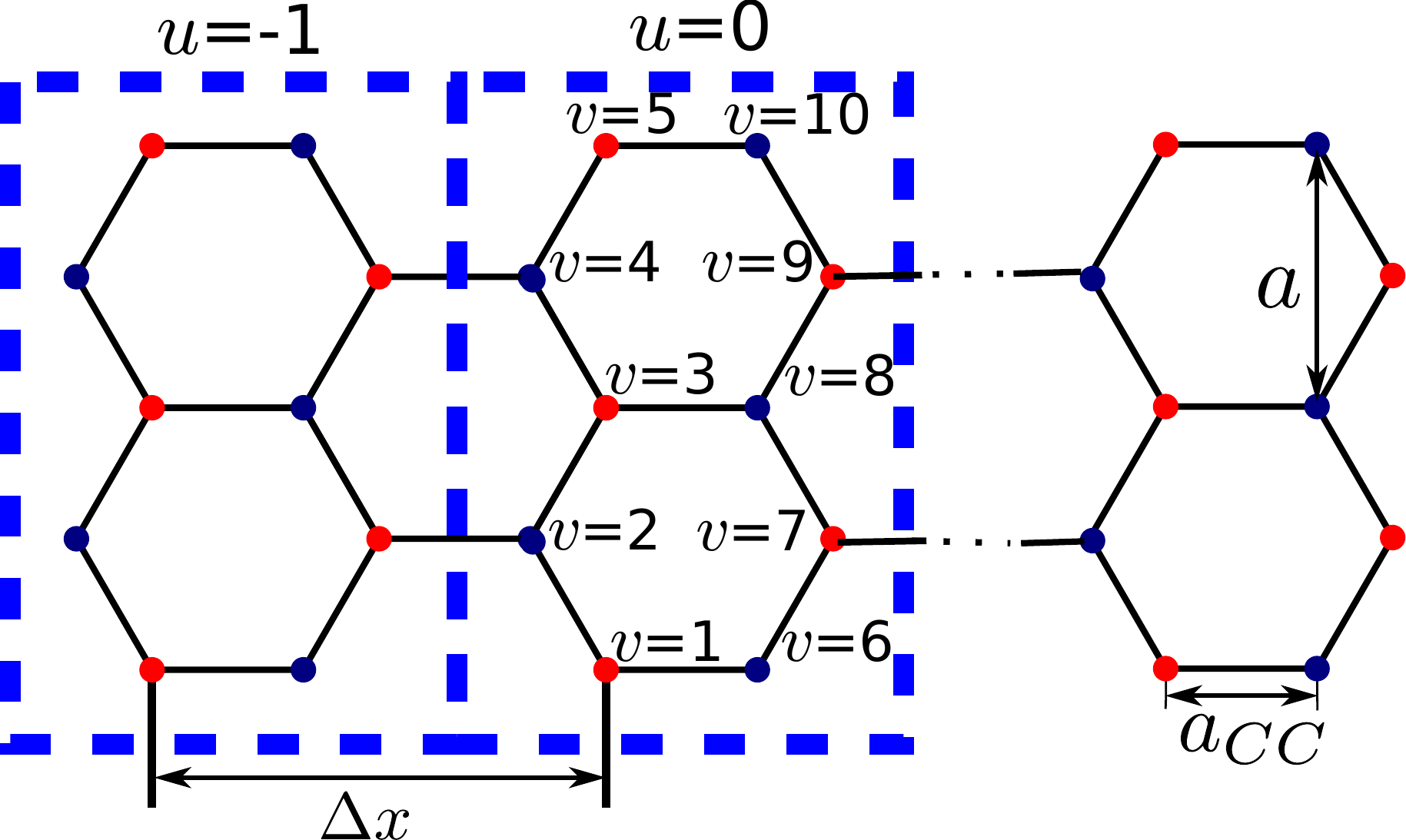}
\caption{ Schematic drawing of the nanoribbons in the input and output channels. In the Bloch waves that serve as the eigenstates of the channels $u$ numbers the elementary cells,
 $v$ the ions within the cell $a_{CC}=0.142$ nm is the nearest neighbor distance, $a=0.246$ nm is the lattice constant, and $\Delta x$ is the period of the considered nanoribbon.}\label{ryszu}
\end{center}
\end{figure}

We use the tight binding Hamiltonian for $\pi$ electrons with the nearest neighbor hopping
\begin{eqnarray}
H=\sum_{\{i,j\}}t_{ij} (c_i^\dagger c_j+c_j^\dagger c_i)+\sum_i V({\bf r}_i) c_i^\dagger c_i, \label{dh}
\end{eqnarray}
and the hopping energy of $t_{ij}=-2.7$ eV.
For the description of the transport at the Fermi level we prepared a
tight-binding  version of the quantum transmitting boundary method \cite{bib1} (for the finite difference
approach see Ref. \cite{bib2}). The studied system consists of an input and output channels and a scattering region.
We assume that the channels are infinite and uniform, i.e. periodic in the atomic scale.
The electron eigenstates in the nanoribbons are determined
in the Bloch form,
\begin{equation}
\psi_{u,v}^k=\chi_v^k e^{iku\Delta x},\label{blw}
\end{equation}
where $k$ is the wave vector, $u$ numbers the elementary cells of the ribbon (see Fig. \ref{ryszu}), $v$ -- the atoms within
the elementary cell, and $\chi_v^k$ is the periodic function, assuming the same
values within each elementary cell of the ribbon. The detailed procedure for determining the dispersion relation and Bloch functions for
the channels is given in the Appendix.

Generally, in the Hamiltonian eigenstates within the channels the scattering wave function is a linear combination of the Bloch functions for a
number of subbands at the Fermi level
\begin{equation}
 \psi_{u,v} = \sum\limits_l ( c_{in}^l \chi_v^{k_+^l} e^{i k_+^l u \Delta x} + d_{in}^{l} \chi_v^{k_-^l} e^{i k_-^l u \Delta x}  ).
\end{equation}
For the electron incident to the scattering region from the left in the input (left) lead far away from the scattering region
the electron wave function takes the form of a superposition of incoming ($k_+$) and backscattered ($k_-$) Bloch waves,
\begin{equation}
 \psi_{0,v} = \sum\limits_l  (c_{in}^l \chi_v^{k_+^l}  + d_{in}^{l} \chi_v^{k_-^l} ) ,
\label{eq:left1}
\end{equation}
where $c_{in}^l$ are the  amplitudes of the incoming states and $d_{in}^{l}$ - of the outgoing ones.
The sum in Eq. (\ref{eq:left1}) runs over the current carrying states
since at a large distance from the scatterers the contribution of the evanescent modes can be neglected.
In the right lead the wave function is a superposition of Bloch waves traveling from the scattering
region, \begin{equation}
 \psi_{N,v} = \sum\limits_l  c_{out}^l \chi_v^{k_+^l} e^{i k_+^l N \Delta x} .
\label{eq:right}
\end{equation}
For solution of the scattering problem we assume that the electron is incident in a single subband $l$,
$c_{in}^l=\delta_{jl}$. Then, the amplitudes $d_{in}^l$ and $c_{out}^l$ are to be determined.
At the boundaries between the leads and the scattering region which we cover within the computational box we require
continuity of both of the wave function and its derivative.
The numerical procedure for determination of the scattering amplitudes is given in the Appendix.

Once the scattering amplitudes are determined we proceed to calculation of transmission $T$ and reflection $R$ probabilities for the $i$th incident mode
according to the Landauer approach \cite{tworzydlo,ccc},
$
\nonumber T_{i} = \sum\limits_j t_{ij}  $ and
$\nonumber R_{i} = \sum\limits_j r_{ij}, $
where $t_{ij}=\left| \frac{c_{out}^{j} }{ c_{in}^{i} } \right|^2 \left| \frac{ (\phi_{k_+})^{j} }{ (\phi_{k_+})^{i} } \right|$ and
$r_{ij}=\left| \frac{ d_{in}^{j} }{ c_{in}^{i} } \right|^2 \left| \frac{ (\phi_{k_-})^{j} }{ (\phi_{k_+})^{i} } \right|$,
are the probabilities of the electron transfer from incident subband $i$ to $j$ subband of the output and input channels, respectively,
and $ (\phi_{k_+})^{i} $ [$(\phi_{k_-})^{i}$] is the probability current flux of the $i$th mode wave function traveling in the right [left] direction.

The currents in the tight binding approach
flow along the $\pi$ interatomic bonds, and the formula for the current flowing from atom $m$ to atom $n$ as derived \cite{waka} from the Schr\"odinger equation
is \begin{equation}
 j_{mn} =\frac{i}{\hbar} \left[ t_{mn} \psi^*_m \psi_n - t_{nm} \psi_n \psi^*_m \right],
\end{equation}
where $\psi_m$ is the wave function at the $m$-th ion.
The flux is evaluated as
\begin{equation}
 \phi = \sum\limits_{l} \sum\limits_{n_l} j_{ln_l},\label{cflux}
\end{equation}
where the sum runs over the atoms across the end of the channel $l$ and their neighbors $n_l$ within the interior of the computational box. 
Finally, the conductance is evaluated from the Landauer formula as $G=\frac{2e^2}{h} \sum_{i}T_i$.
Since the formula involves sum over incident subbands, for the calculations it is sufficient to consider the solution of the Schr\"odinger equation for a single input channel 
only. Note, that our choice of the quantum transparent boundary conditions implies working with the wave functions -- which is an alternative to the
Greens function method \cite{abnormal,wimmer}, that deals with Hamiltonian operators mainly. The choice between the two is a matter of taste.

The effective potential of the charged tip as seen by the Fermi level electrons is a result of the screening of its Coulomb potential
by the electron gas. The effective potential as obtained by the Sch\"odinger-Poisson modeling \cite{kmk} can be approximated
by a Lorentz function
\begin{equation}
V(x,y)=\frac{V_t}{1+\left( (x-x_t)^2+(y-y_t)^2\right)/d^2},
\end{equation}
where $x_t,y_t$ are the coordinates of the tip position, $d$ -- the width of the effective tip potential and $V_t$ -- its height.
The height of the tip potential is determined by the charge accumulated by the tip, and $d$ is of the range of the tip - electron gas distance \cite{kmk}.
A discussion of the results for varied $V_t$ and $d$ parameters is provided below.

\section{Conductance mapping for nanoribbons}
We consider first the zigzag ribbons - labeled by ZZ in the Figures, and next proceed to armchair: semiconducting (AS) and metallic (AM) ribbons.
\subsection{Zigzag ribbons}

\begin{figure*}[htbp]
 \includegraphics[width=0.2\textwidth]{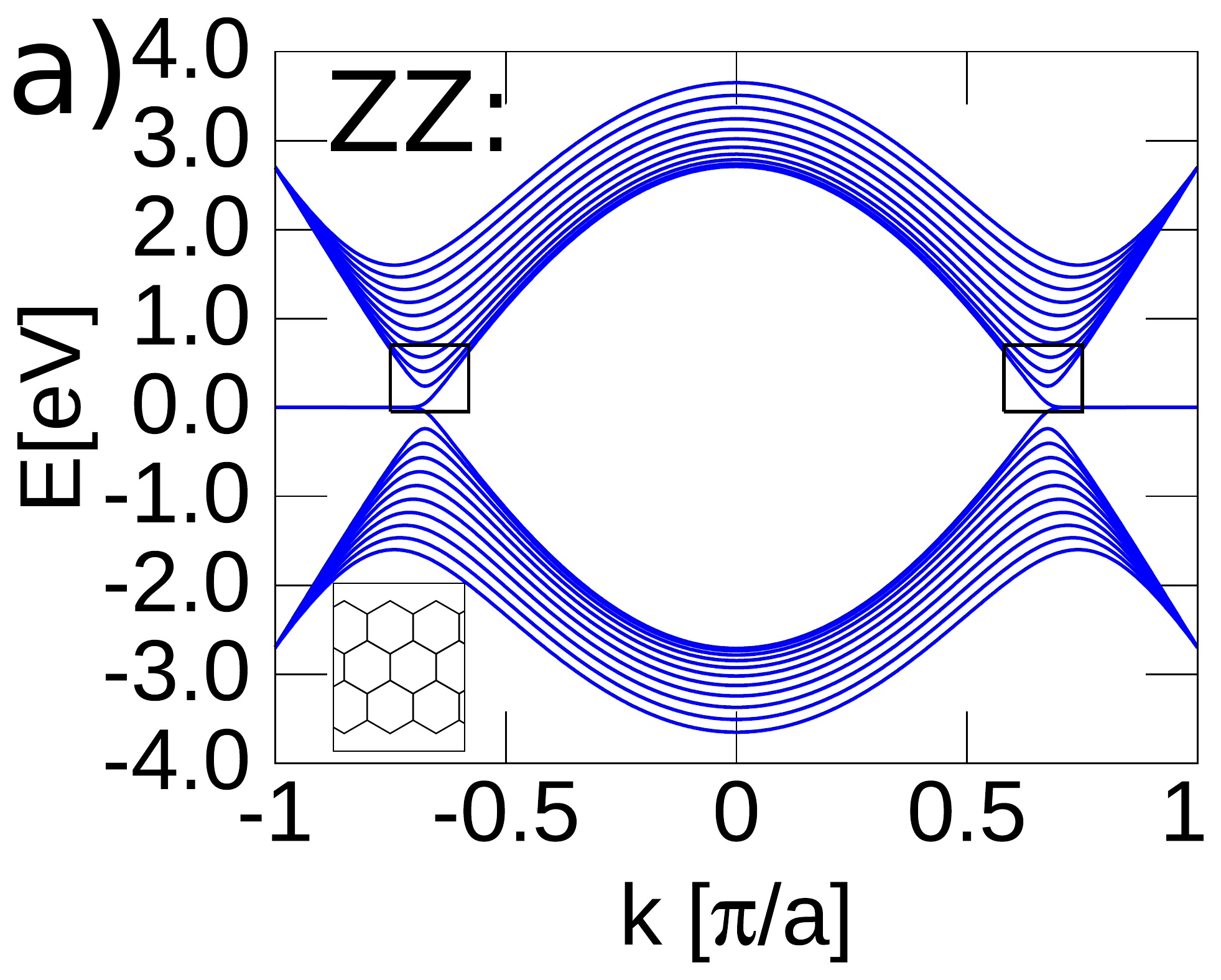}
 \includegraphics[width=0.2\textwidth]{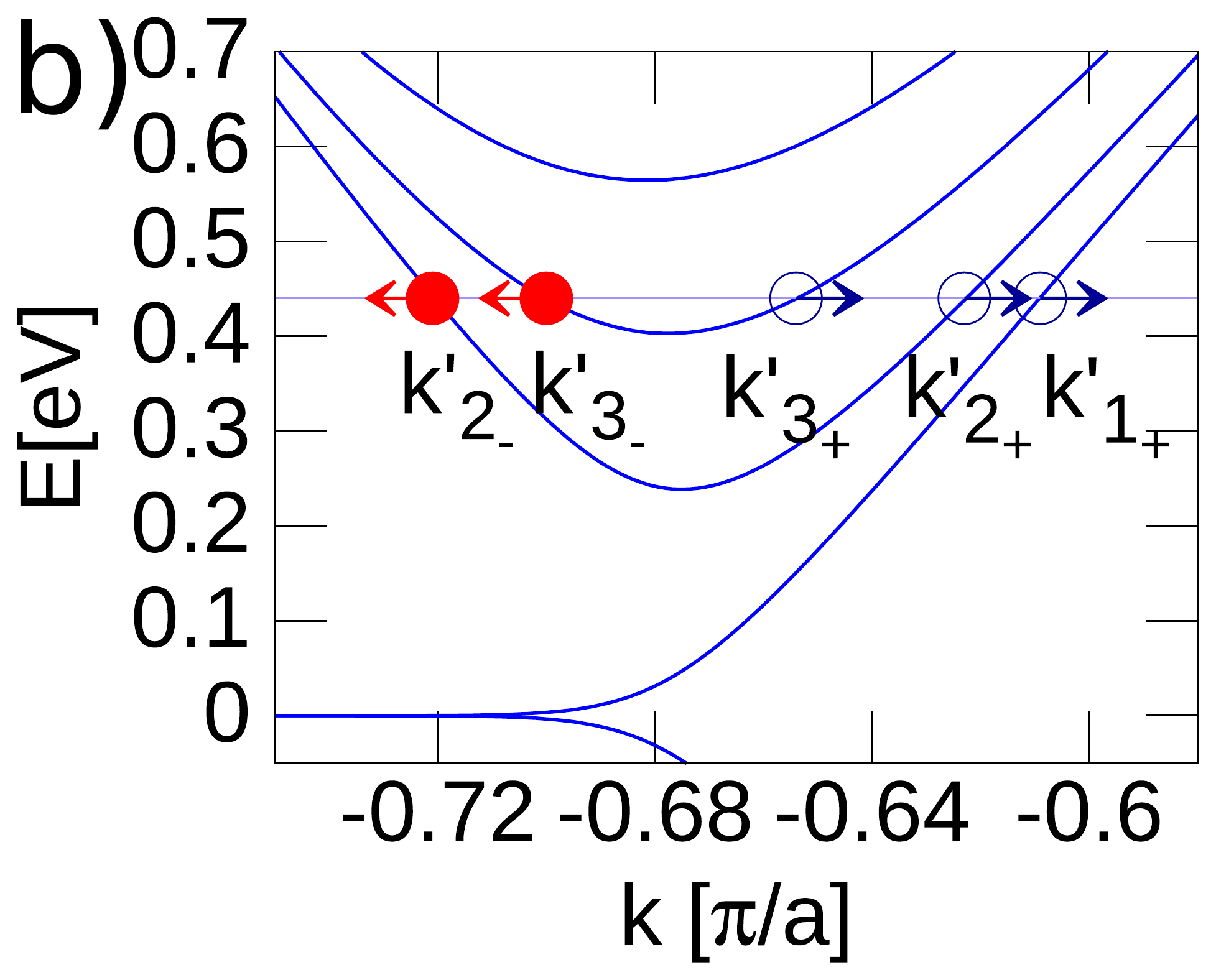}
 \includegraphics[width=0.2\textwidth]{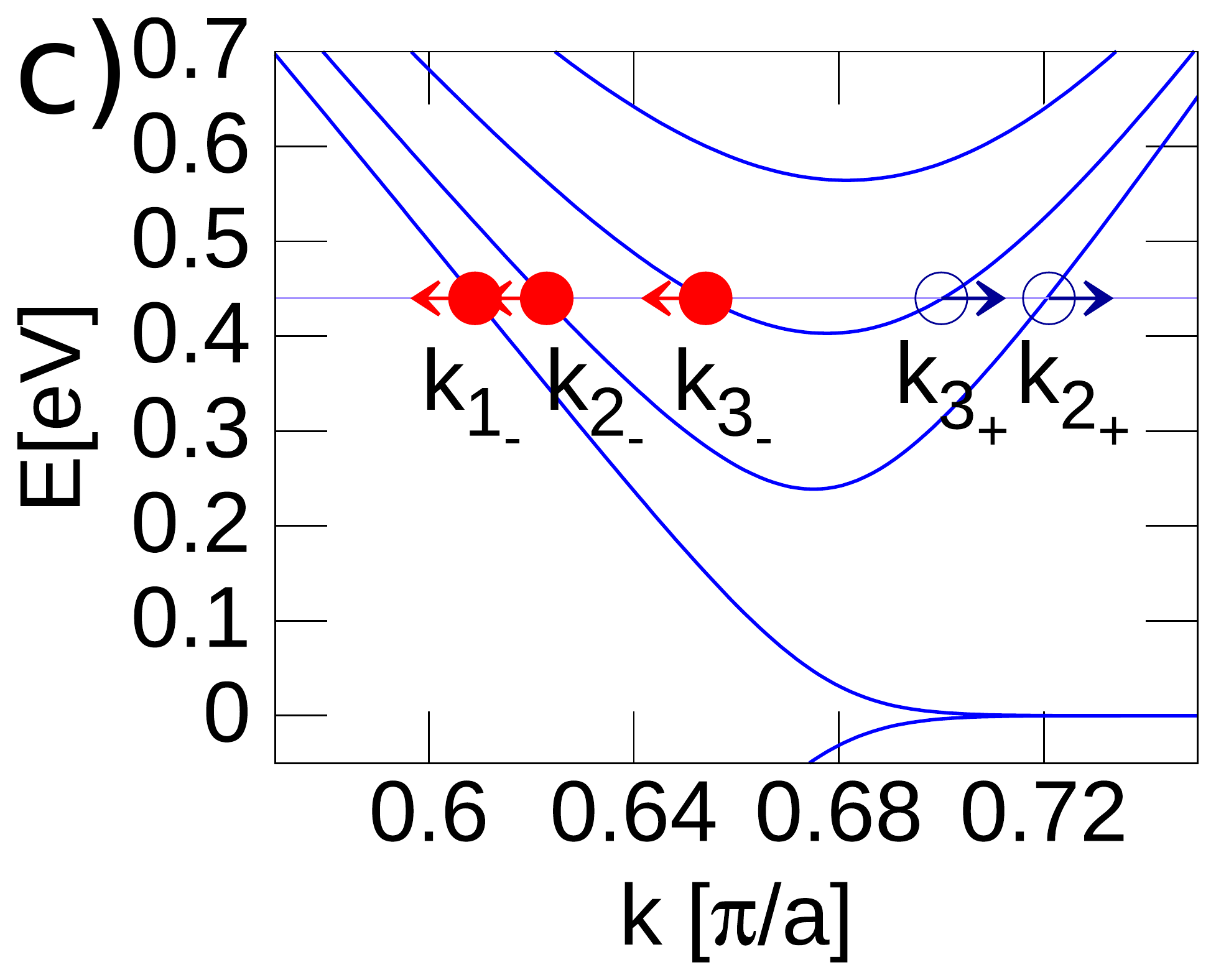}\\
 \includegraphics[width=0.2\textwidth]{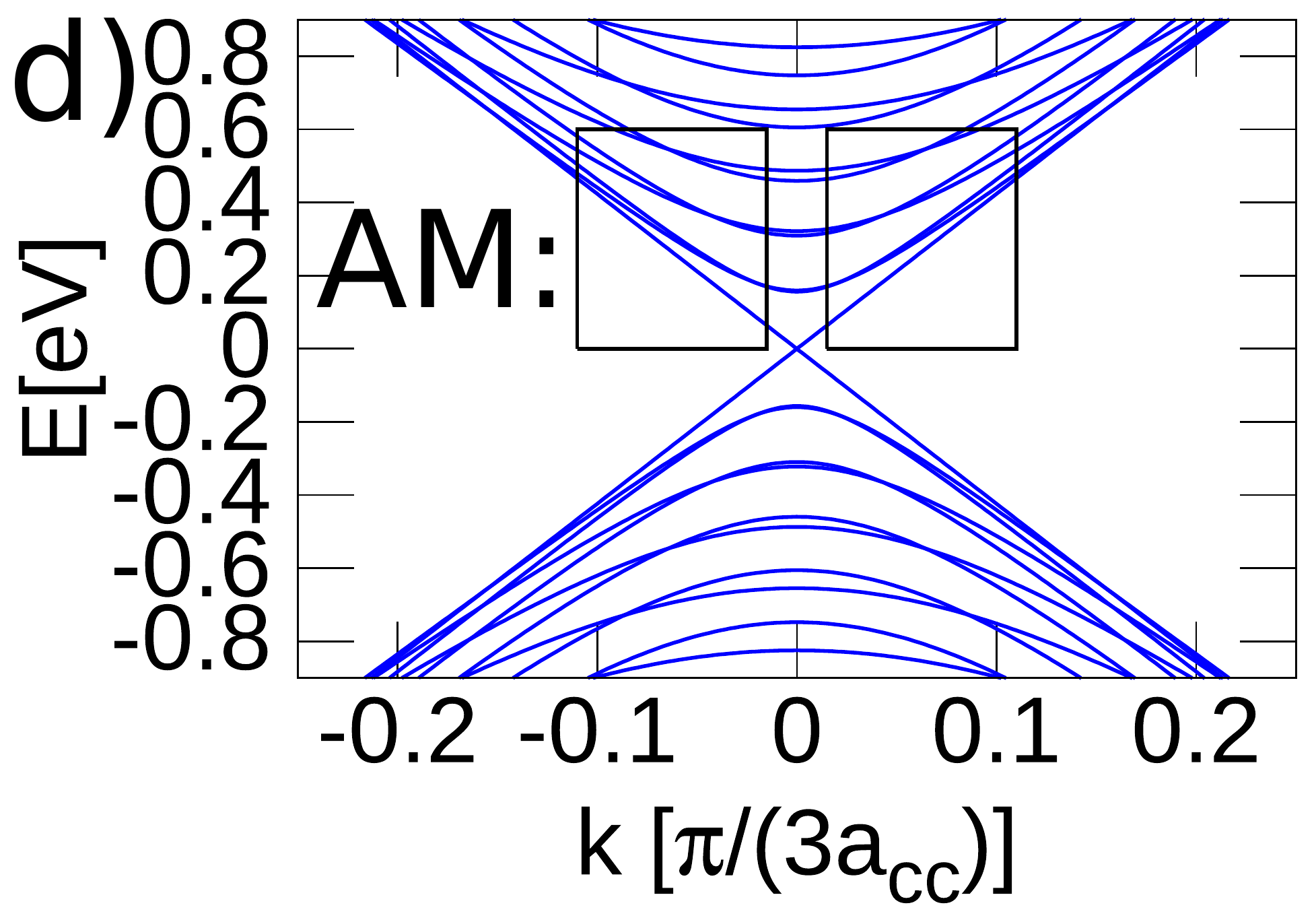}
 \includegraphics[width=0.2\textwidth]{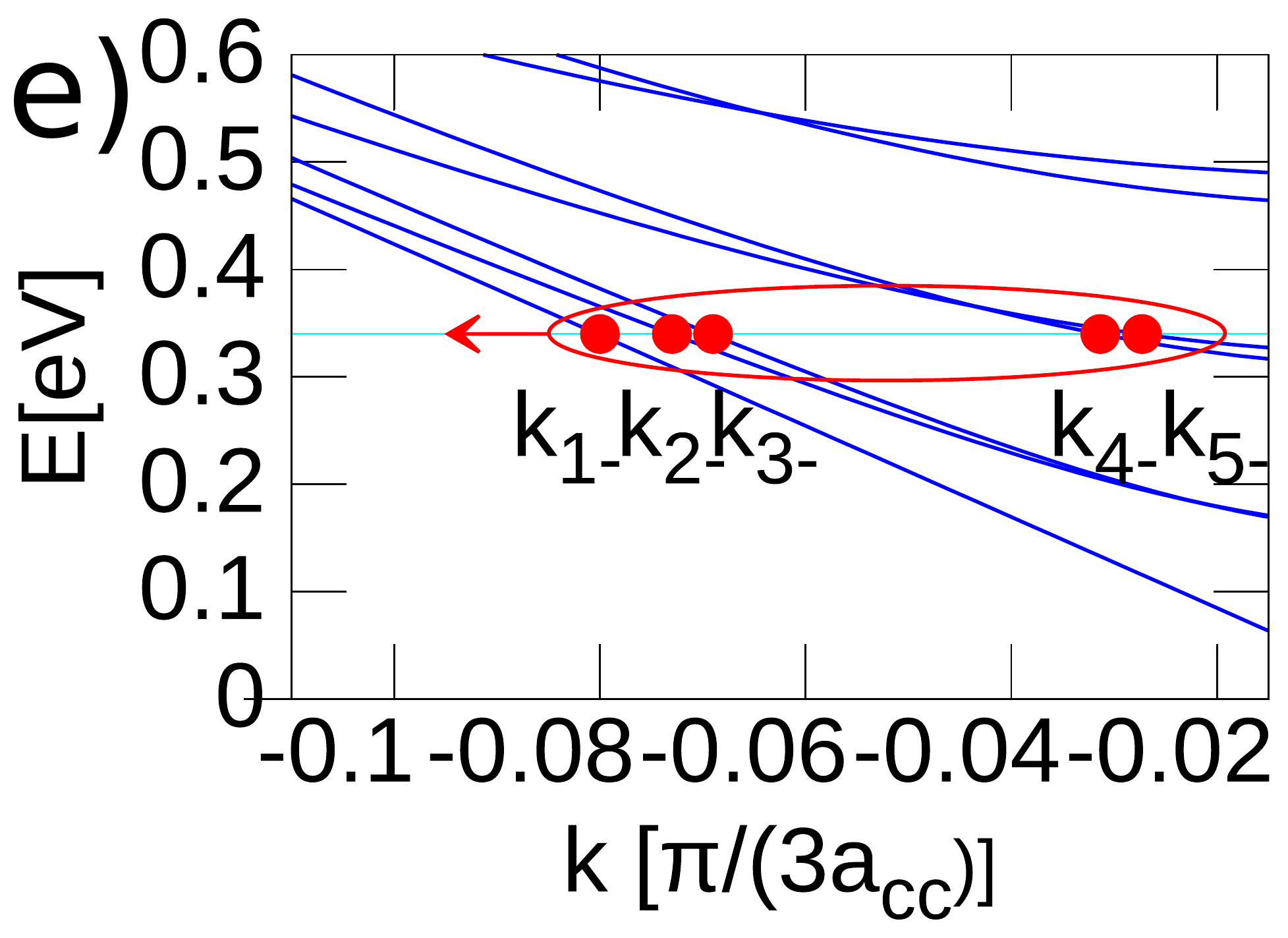}
 \includegraphics[width=0.2\textwidth]{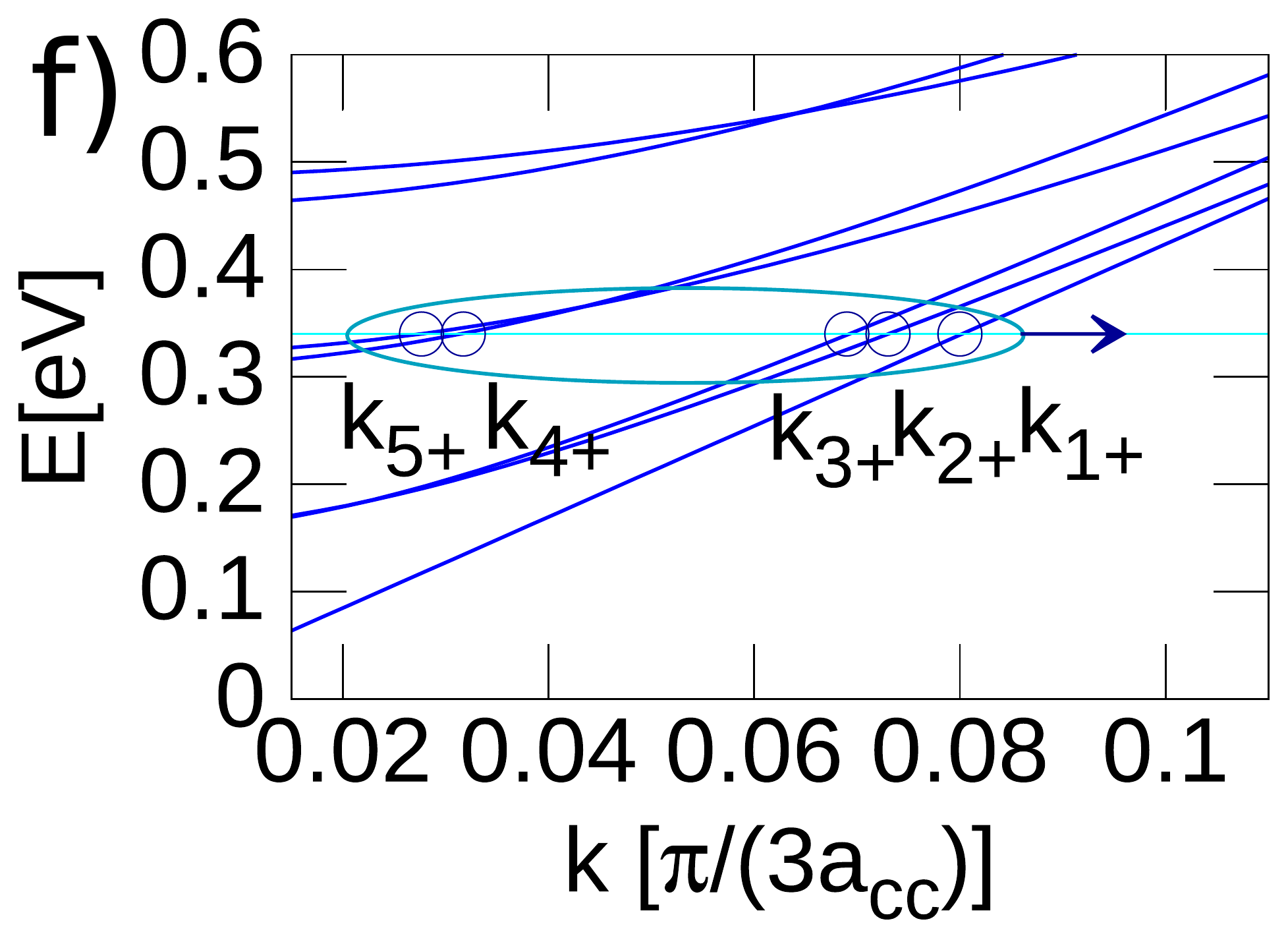}\\
 \includegraphics[width=0.2\textwidth]{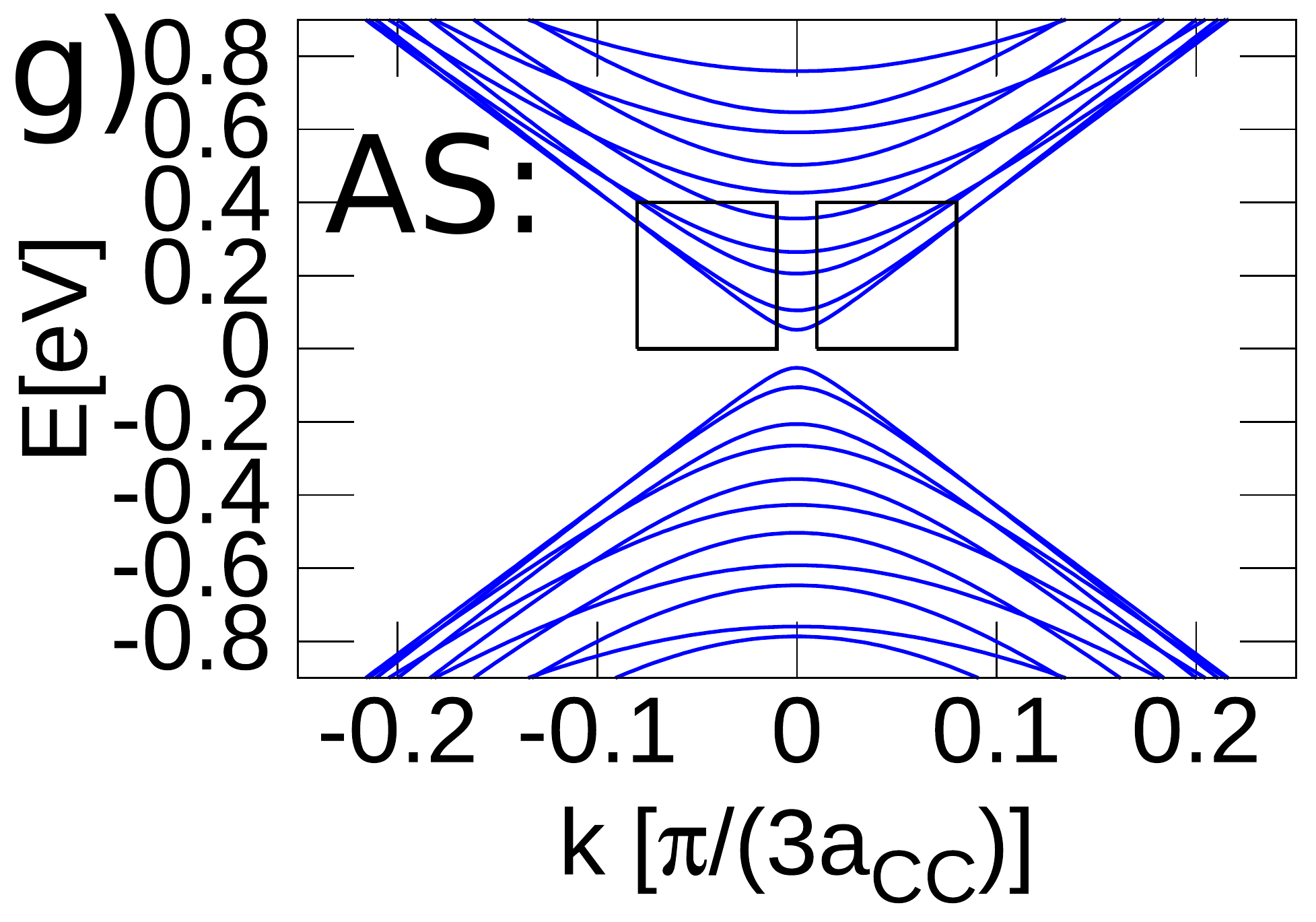}
 \includegraphics[width=0.2\textwidth]{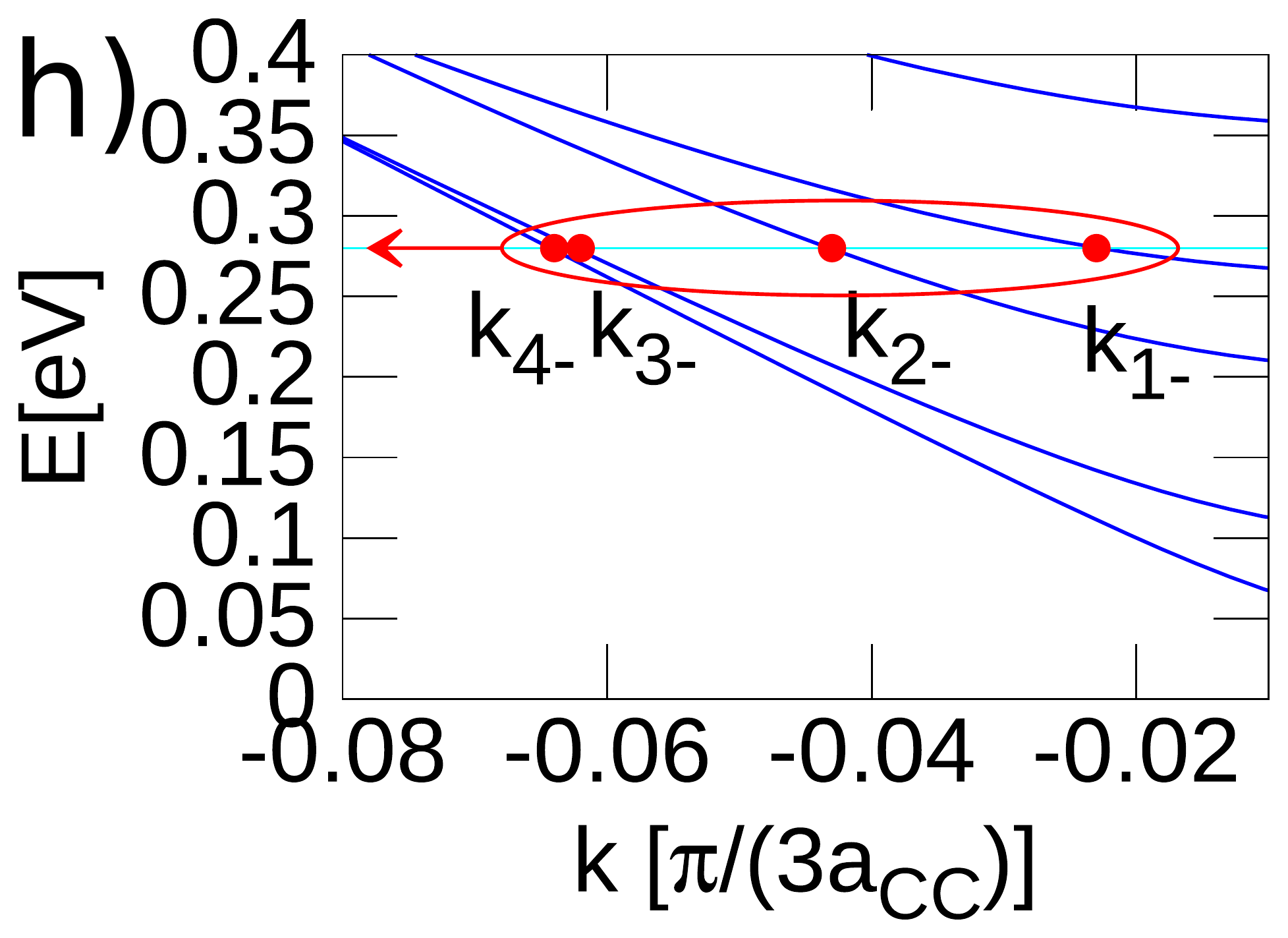}
 \includegraphics[width=0.2\textwidth]{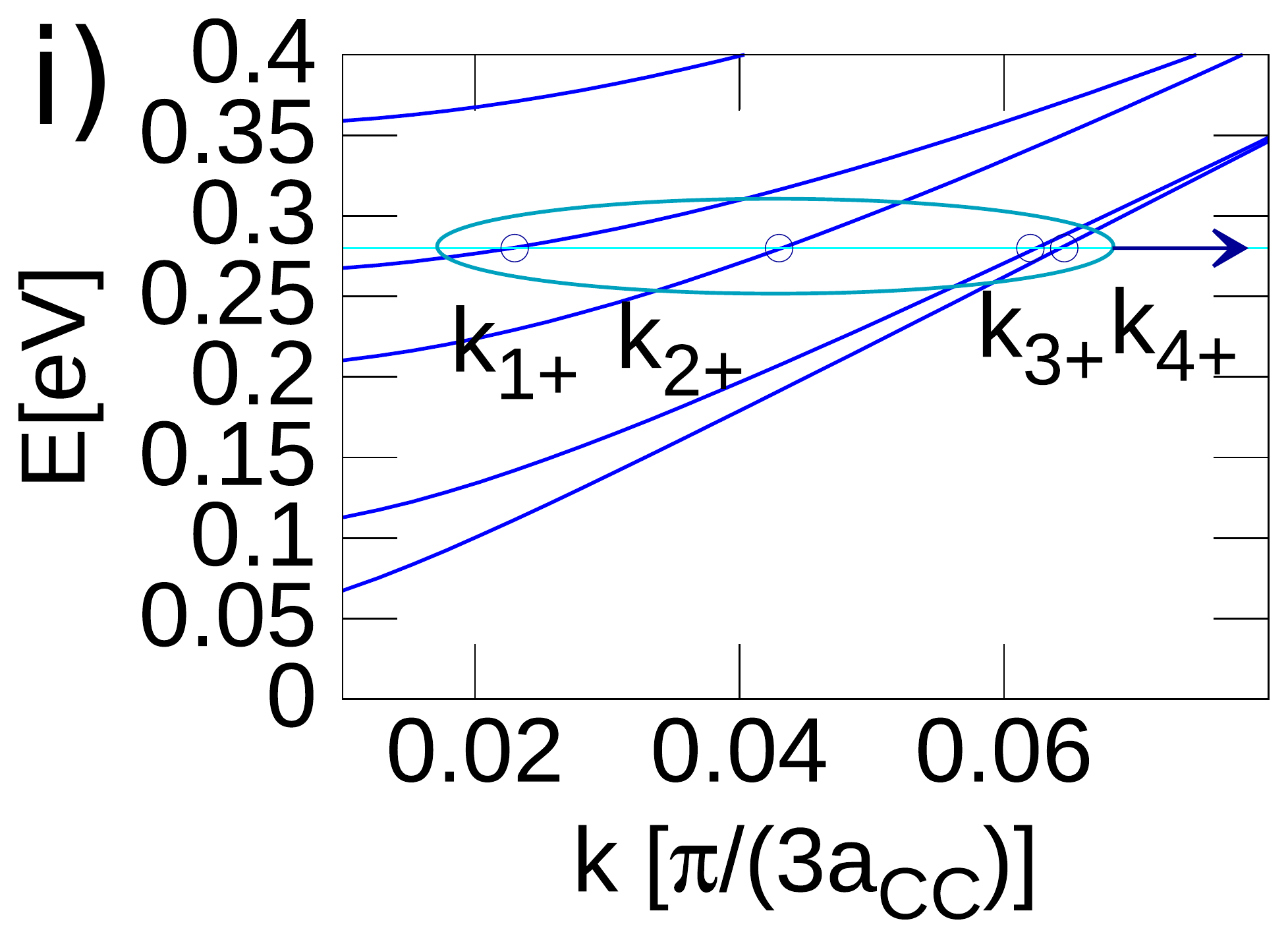}

\caption{(a) Dispersion relation for a zigzag nanoribbon 102 atoms wide.
(b,c) Enlarged fragments marked by rectangles in (a). The arrows indicate the
current direction.
(d-f) same as (a-c) only for a metallic  armchair ribbon with 92 atoms across the channel.
(g-h) same as (a-c) only for a semiconducting armchair ribbon with 93 atoms across the channel.
Arrows in (b,c,e,f,g,h) indicate the direction of the current flow.
Here and in the other figures ZZ, AM, and AS stand for zigzag, armchair metallic and armchair semiconducting ribbons.
} \label{disprel}  
\end{figure*}

Figure \ref{disprel}(a) shows the dispersion relation for a zigzag nanoribbon
with 102 atoms across the channel.
The zigzag edge does not couple the states of the $K$ and $K'$ valleys of graphene which are thus present in the dispersion relation ($K,K'=\pm \frac{2\pi}{3 a}$).
The dispersion relation is given in Figs. \ref{disprel}(a-c). For Fermi energy $E_F=0.44$ eV
  there are five eigenstates with current flowing to the right ($k_{1'_+},k_{2'_+},k_{3'_+},k_{2_+}, k_{3_+}$) and to the left ($k_{1_-},k_{2'_-},k_{3'_-},k_{2_-}, k_{3_-}$).

\begin{figure*}[htbp]
\begin{tabular}{ccc}
 \includegraphics[width=0.2\textwidth]{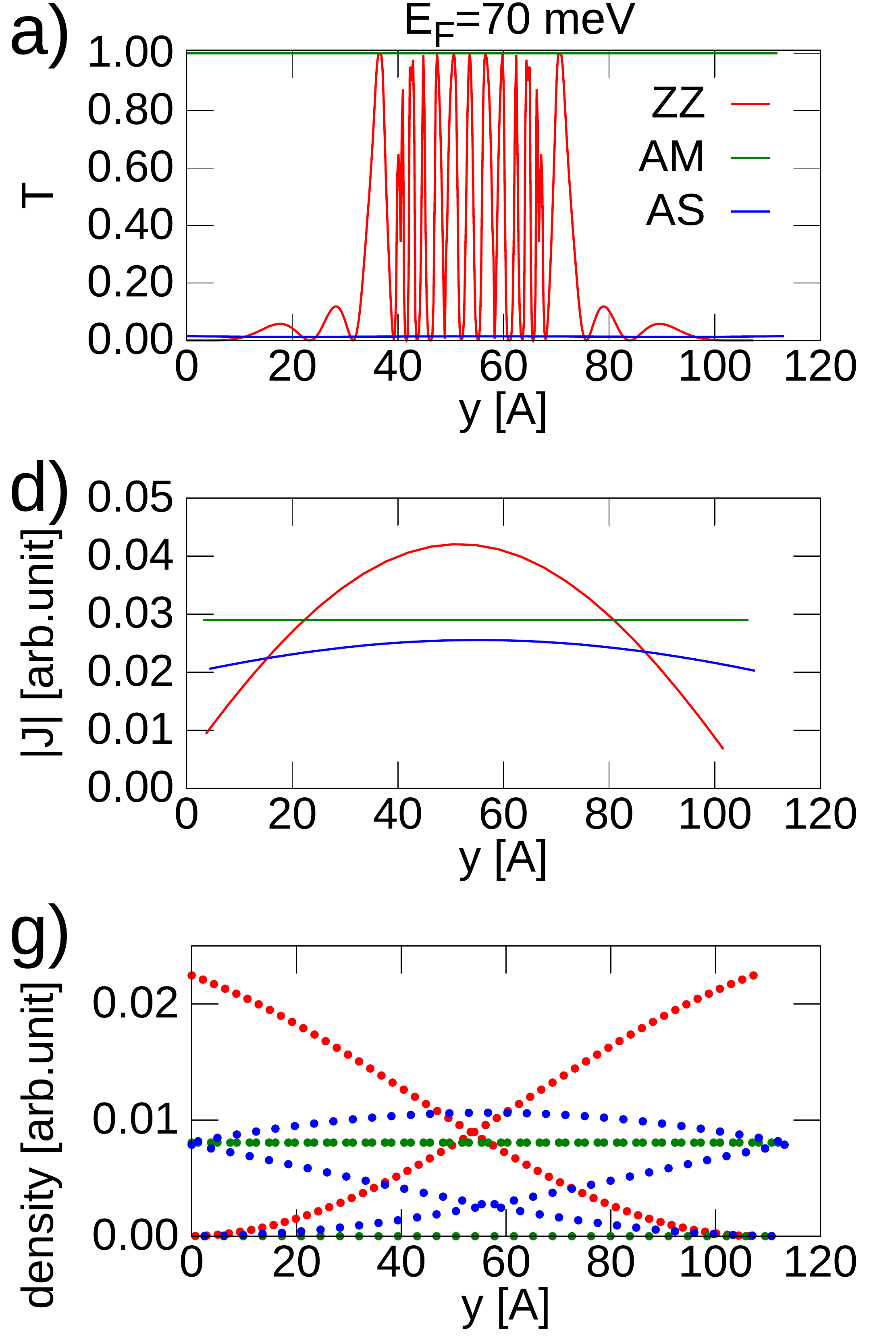}&
 \includegraphics[width=0.2\textwidth]{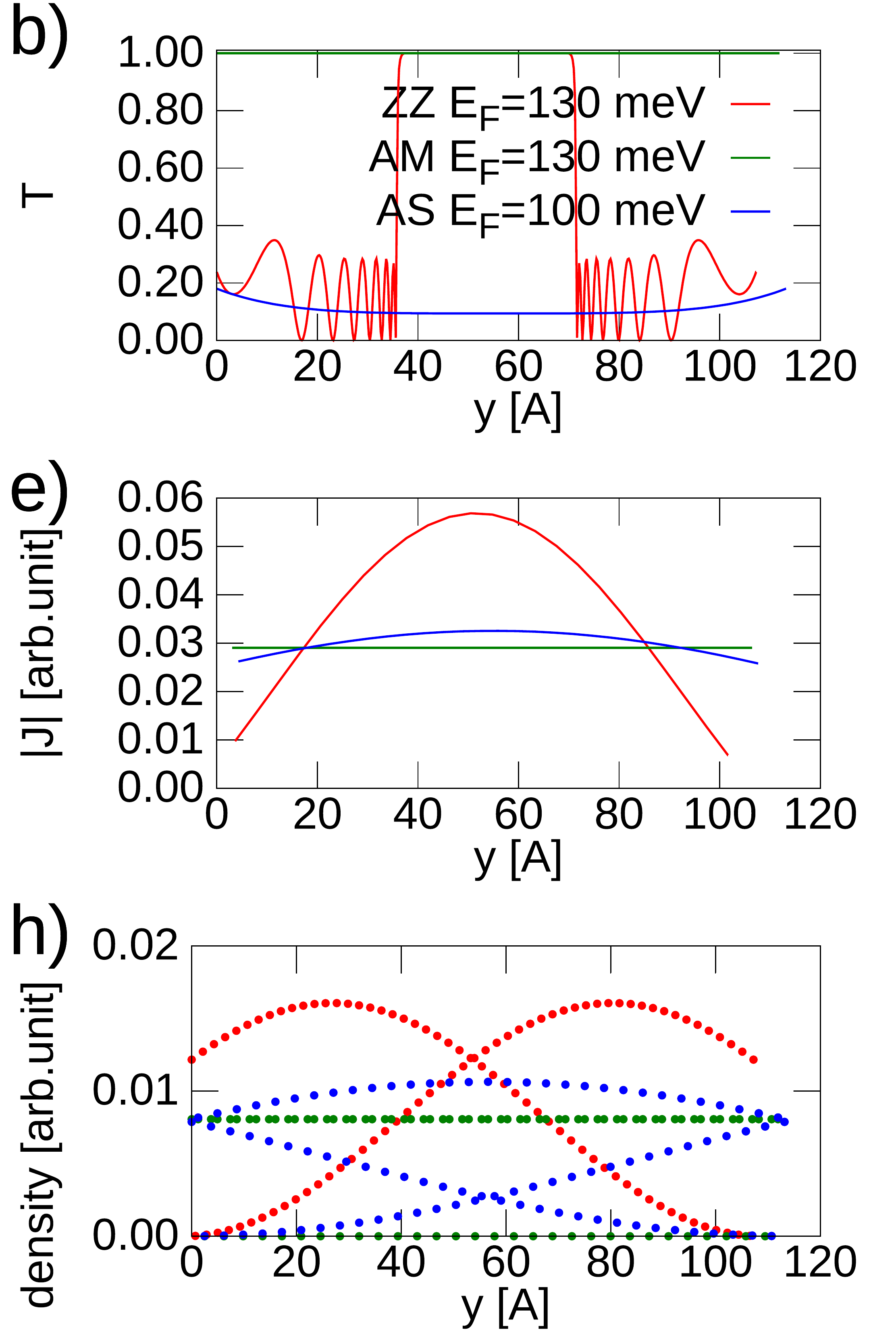} &
  \includegraphics[width=0.2\textwidth]{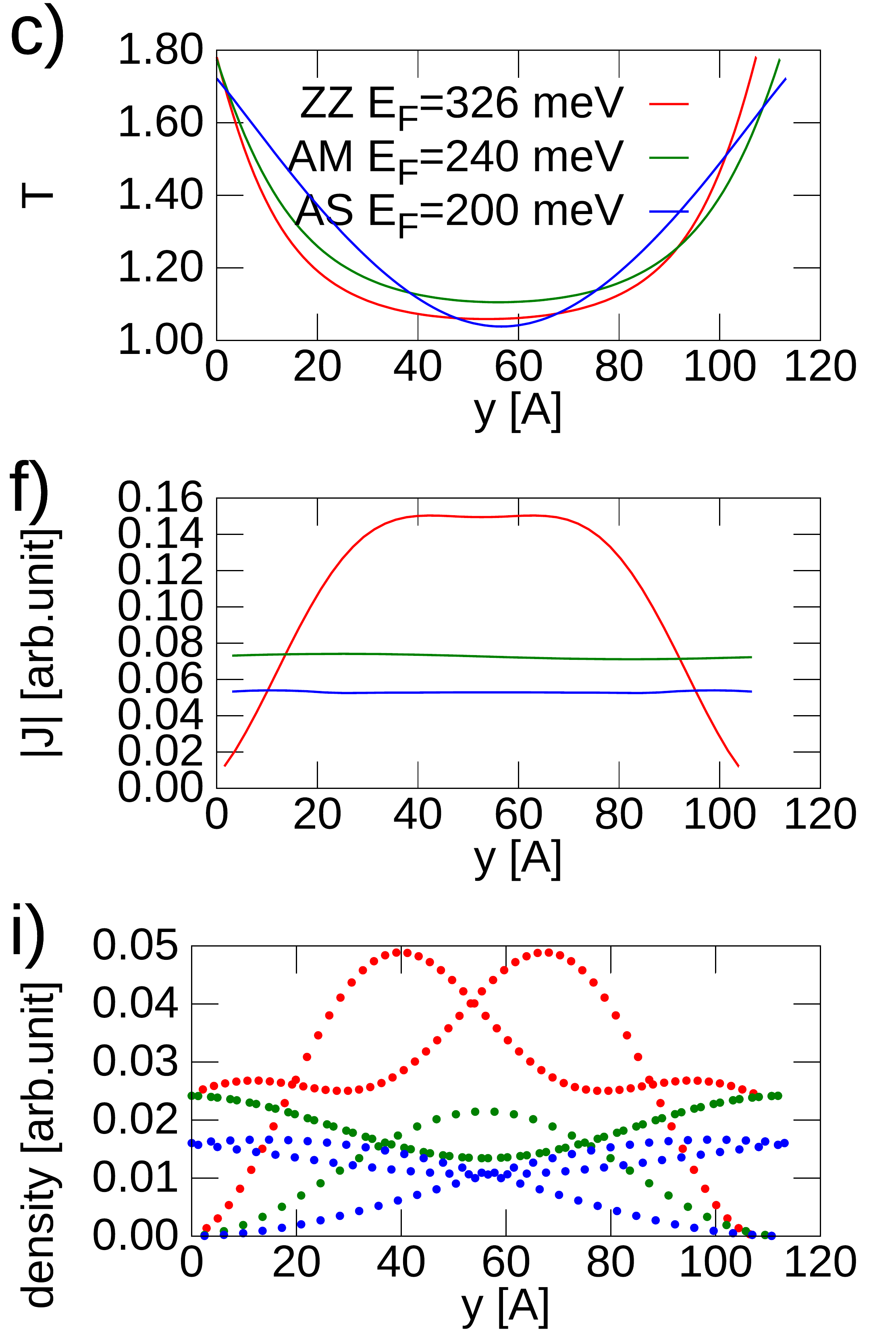} \\
 \end{tabular}
\caption{
(a-c) The transfer probability as a function of the tip position for $V_t=0.2$ eV and $d=40a$.
(d-f) The density current averaged over neighbor cells.
(g-i) The probability density for ribbons without the tip.
In the left column (a,d,g) $E_F=0.7$ eV and in the middle (b,e) $E_F=0.13$ eV for zigzag 
and armchair metallic and (h) $E_F=0.1$ eV for armchair semiconducting.
Red lines correspond to a zigzag ribbon with 102 atoms across the channel, green (blue)
to the metallic (semiconducting) armchair ribbons with 92 (93) atoms across the channel.
In (c,f,i) results for higher energies are given.
For the corresponding dispersion relation see Fig. \ref{disprel}(a-c) for the zigzag ribbon, and Fig. \ref{disprel}(d-f) for metallic and Fig. \ref{disprel}(g-i) for semiconducting nanoribbons.}\label{lowestsb}
\end{figure*}

\begin{figure*}[htbp]
 \includegraphics[width=0.8\textwidth]{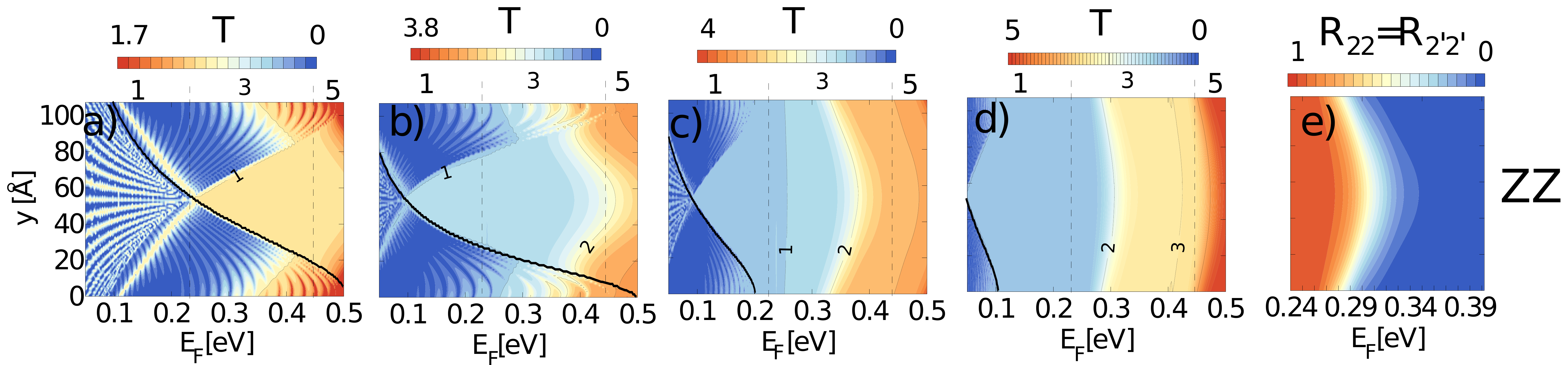}
\caption{ Conductance of the zigzag ribbon with dispersion relation of Fig. \ref{disprel} as a function
of the position of the tip and the Fermi energy. Parameters of the tip $V_t=0.5$ eV, $d=40a$ (a), $V_t=0.5$ eV, $d=20a$ (b),
$V_t=0.2$ eV, $d=40a$ (c), $V_t=0.1$ eV, $d=40a$ (d-f). The dashed lines separate regions of
a varied number of subbands at the Fermi level -- given above the upper axis.
The thick black line in (a-d) corresponds to the tip positions for which the potential at the edge is raised to the Fermi energy.}\label{zigzagtip}
\end{figure*}

\begin{figure}[htbp]
 \includegraphics[width=.4\textwidth]{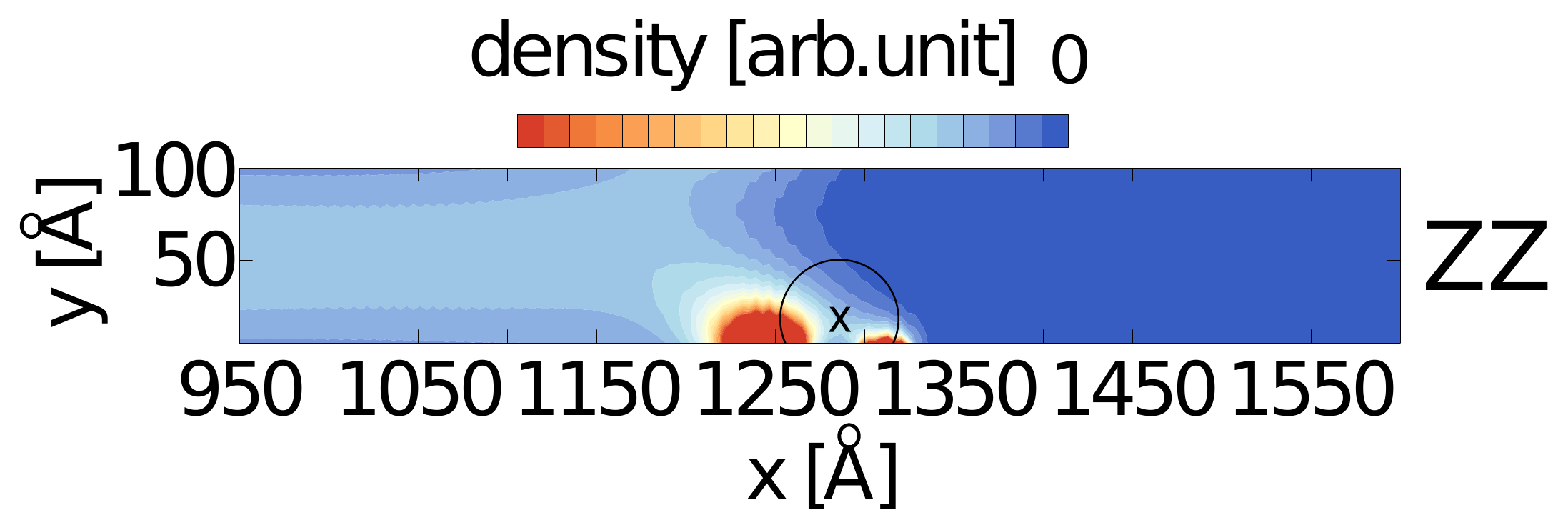}
\caption{Scattering probability density for $V_t=0.2$ eV and $d=40a$ at $E_F=130$ meV and the tip near the lower edge of the ribbon (cross). The circle
corresponds to potential introduced by the tip equal to $E_F$.} \label{sztip}
\end{figure}

\begin{figure}[htbp]
 \includegraphics[width=0.5\textwidth]{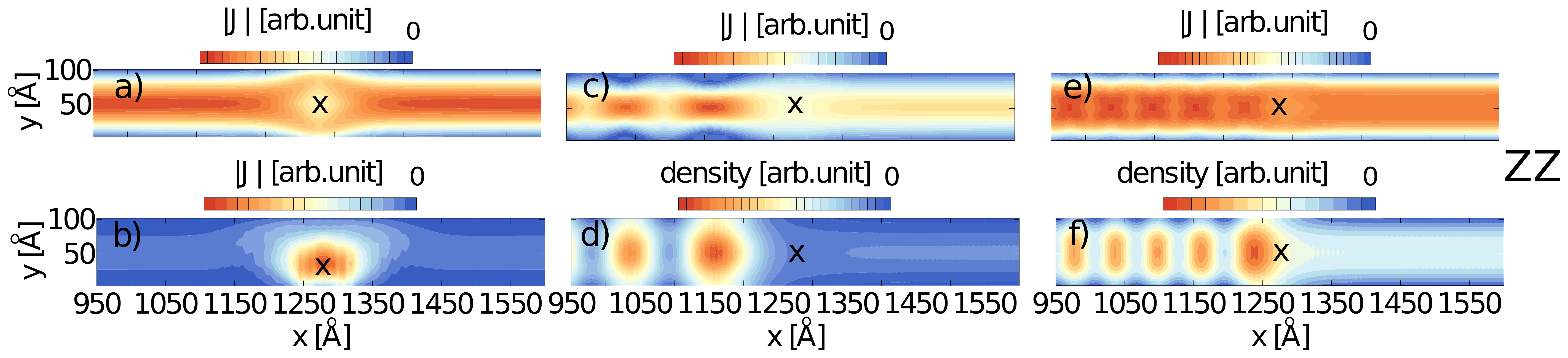}
\caption{(a,b) Amplitude of the current for the zigzag ribbon for $V_t=0.2$ eV, $d=40a$ for energy $E_F=130$ meV in the lowest subband transport
and the tip in the center of the ribbon (a) -- for which $T=1$ and near the lower edge (b) of the ribbon [$T=0$]. (c-f) results for three subbands at the Fermi level
for the amplitude of the current (c,e) and the scattering density (d,f) for $E_F=0.3$ eV and $E_F=0.4$ eV. The summed transfer probability is 1 at (c,d) and 3 at (e,f).}
\label{prondy}
\end{figure}

Let us consider the ribbons conductance response to the tip potential for a low energy
of the incident electron.
Figures \ref{lowestsb}(a-b) show the transfer probability as a function of the tip position across the ribbon for two values of $E_F$
corresponding to the lowest subband transport (electron incoming with wave vector $k_{1'_+}$, that can be backscattered only to $k_{1_-}$ -- see Fig. \ref{disprel}(b,c)).
For both the considered Fermi energies the conductance gets low when the tip is located near the edge, in spite of the fact
that the current for the unperturbed zigzag ribbon vanishes at the edges [Figure \ref{lowestsb}(d-e)]: at the zigzag edges only the ions of a single sublattice are  occupied in the Hamiltonian eigenstates [Figure \ref{lowestsb}(g-h)]
and the current flows between the ions belonging to different sublattices \cite{nrb4}.
For the central position of the tip we have either $T=1$ for larger $E_F=0.13$ eV [Fig. \ref{lowestsb}(b)] or a rapid variation of $T$ between 0 and 1
for the lower $E_F=0.07$ eV [Fig. \ref{lowestsb}(a)].
In the plot of $G$ as a function of the tip position and the Fermi energy given in Fig. \ref{zigzagtip}
we find a wedge-shaped flat region with $T=1$.
This region is related to the perfectly conducting channel \cite{pcc,pcc2} in the zigzag ribbon.
Perturbation by the tip -- even if large -- does not induce scattering between the $K$ and $K'$ valleys
as long as the tip potential is slowly varying at the atomic scale.
 Since the tip potential has a long range character there is no intervalley scattering, and in consequence -- in the lowest subband transport conditions -- the backscattering is absent,
 unless the tip is located near the edge of the ribbon [cf. Fig. \ref{sztip}].
The tip -- placed in the center of the ribbon -- is not exactly transparent for the
electron flow,  since the current makes its way between
the edge [Figure \ref{prondy}(a)], but anyway no conductance response is observed.
  For the tip near the edge,
the current circulates around the defect before it is backscattered (see Fig. \ref{prondy}(b) -- the illustrated case corresponds to $T=0$).
 The edge mediates in the intervalley scattering provided
that its potential is raised to the Fermi energy \cite{wurm} by e.g. the tip
inducing formation of a local n-p junction.
 In Fig. \ref{zigzagtip} we marked with the
black line the position of the tip for which its potential at the lower ribbon edge is raised to the Fermi energy,
with a perfect agreement with the boundaries of $T=1$ wedge region.
The transfer probability as a function of the tip position
plotted in Fig. \ref{lowestsb}(a) for low $E_F$ exhibits a rapid variation between 0 and  1 which corresponds to the case when the tip potential
at {\it both} edges exceeds $E_F$, hence multiple intervalley scattering events occur, resulting in either $T=0$ or $T=1$.

The $T=1$ wedge region continues smoothly into the regime of  a few transport subbands at the Fermi level [see Fig. \ref{zigzagtip}].
For larger energies, in the right ($K$) valley of the dispersion relation [Fig. \ref{disprel}(a,c)] the number of modes carrying the current to the right is by 1 larger than the number of modes carrying the current
to the left [Fig. \ref{disprel}(c)]. Hence, even for a strong but long-range perturbation the ribbon carries current at least in a single mode \cite{pcc,pcc2} unless
the tip raises the edge potential to the Fermi level -- as it is still the case in Fig. \ref{zigzagtip}(a,b) also beyond the single subbands transport conditions.

\begin{figure}[htbp]
 \includegraphics[width=0.5\textwidth]{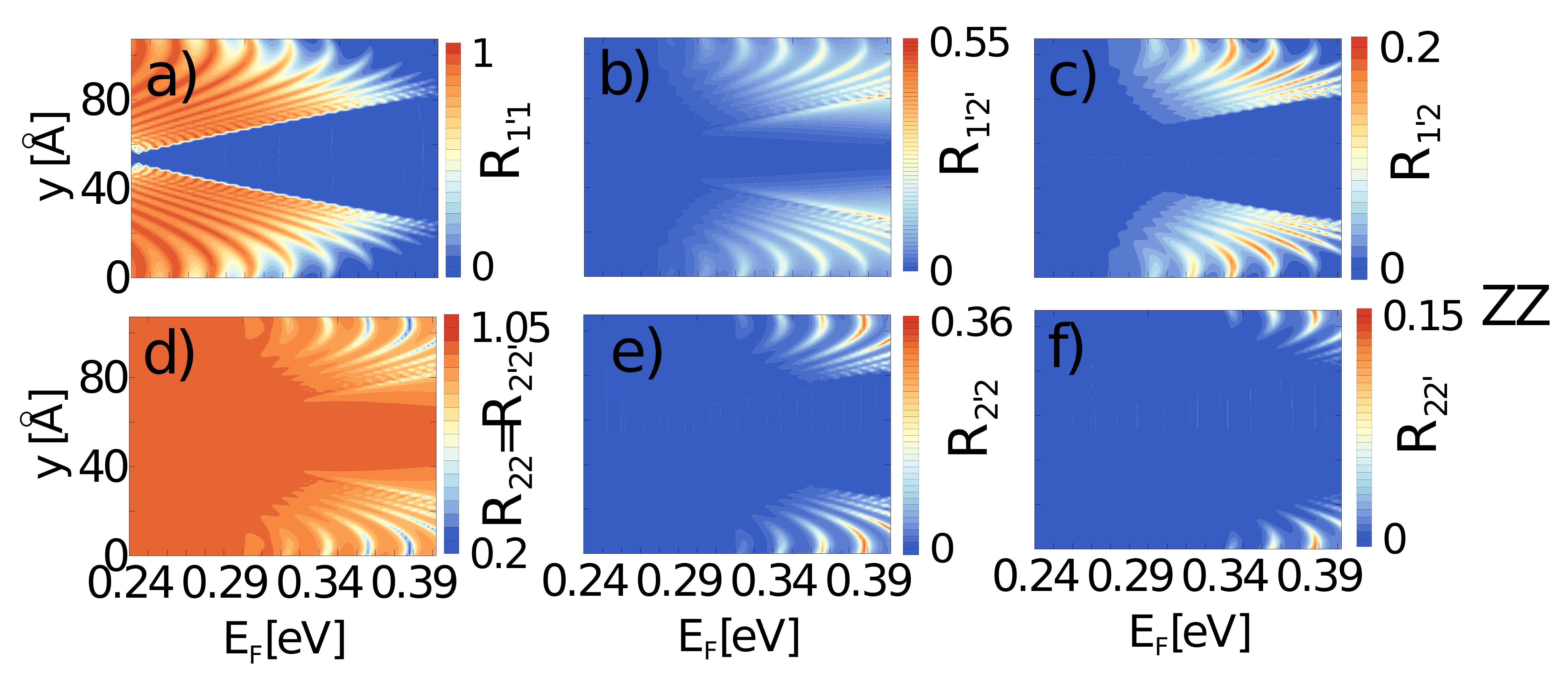}
\caption{ Intersubband backscattering probabilities for parameters of Fig. \ref{zigzagtip}(a)  a) $R_{1'1}$, b) $R_{1'2'}$, c) $R_{1'2}$, d) $R_{22}$, e) $R_{2'2}$, f) $R_{22'}$ for $V_t=0.5$ eV, $d=40a$. }\label{zzsb}
\end{figure}

In Figure \ref{zzsb} we plotted the backscattering
probabilities for the parameters of the tip of Fig. \ref{zigzagtip}(a) (large and wide tip potential $V=0.5$ eV, $d=40$ nm) for the $E_F$ range with three subbands or six wave vectors at the Fermi energy [Fig. \ref{disprel}(b-c)]. The intervalley scattering in the lowest subband (of probability $R_{1'1}$ corresponding to the change of the wave vector from $k'_{1+}$ to $k_{1-}$ -- see Fig. \ref{disprel}(b,c))  occurs only for the tip near the edge of the ribbon [Fig. \ref{zzsb}(a)],
but when it does, it is nearly complete  ($R_{1'1}\simeq 1$). The other intervalley scattering paths $R_{1'2}$ [Fig. \ref{zzsb}(c)], $R_{2'2}$ [Fig. \ref{zzsb}(e)],
$R_{22'}$ [Fig. \ref{zzsb}(f)] occur also for the tip near the edge but with smaller probabilities.
The effective backscattering channels for the tip in the center of the ribbon  are  $R_{2'2'}=R_{22}$  [Fig. \ref{zzsb}(d)] that appear within the same parabolic subbands [Fig. \ref{disprel}(b,c)].
The backscattering by the tip
in the $k'_{2+}\rightarrow k'_{2-}$  channel governs the current and scattering density pattern
that is displayed in Fig. \ref{prondy}(c-f). The oscillations at the left to the tip are due to the interference between $k_{2'_+}$ and $k_{2'_-}$
eigenstates and correspond to the period of $ \frac{2\pi}{|k_{2'_+}-k_{2'_-}|}$  , which is equal to $103 \AA$ and $58 \AA$
for $E_F=0.3$ and 0.4 eV, respectively.
The conductance of the ribbon raises from $1$ to $3$ for $E_F$ above 0.2 eV [see the dashed lines in Fig. \ref{zigzagtip}]
with backscattering by the tip in the $2'_+\rightarrow 2'_-$ channel disappearing at higher Fermi energies,
[cf. Fig. \ref{zigzagtip}(e)].

\subsection{Armchair ribbons}

\begin{figure}[htbp]
 \includegraphics[width=0.35\textwidth]{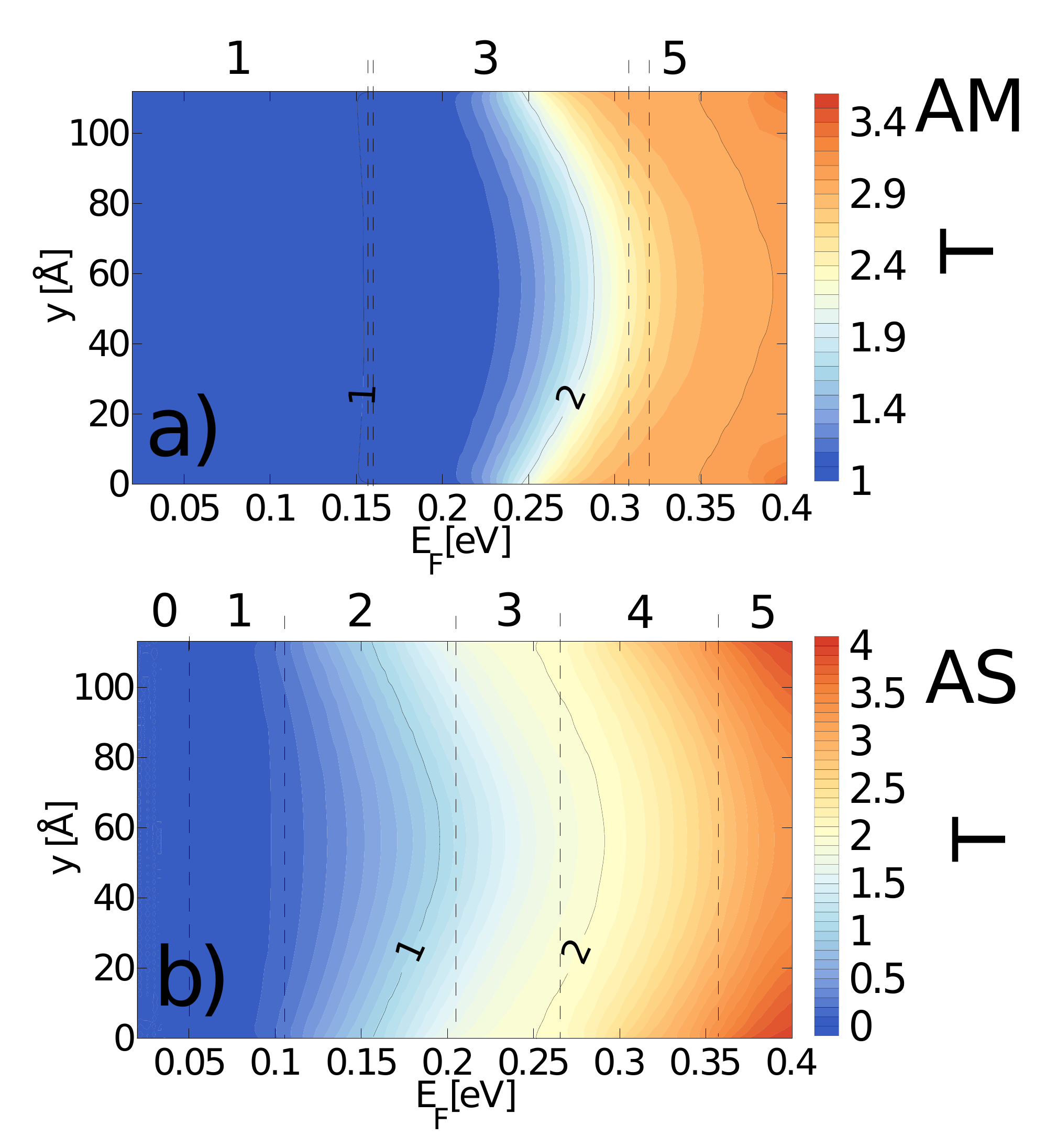}
\caption{Conductance versus the $E_F$ and tip position for a metallic (a) and semiconducting (b) nanoribbons for $V_t=0.2$ eV, and $d=40a$. The dashed lines separate the regions with varied number of subbands at the Fermi level.
 } \label{armchairskan}
\end{figure}

\begin{figure}[htbp]
 \includegraphics[width=0.46\textwidth]{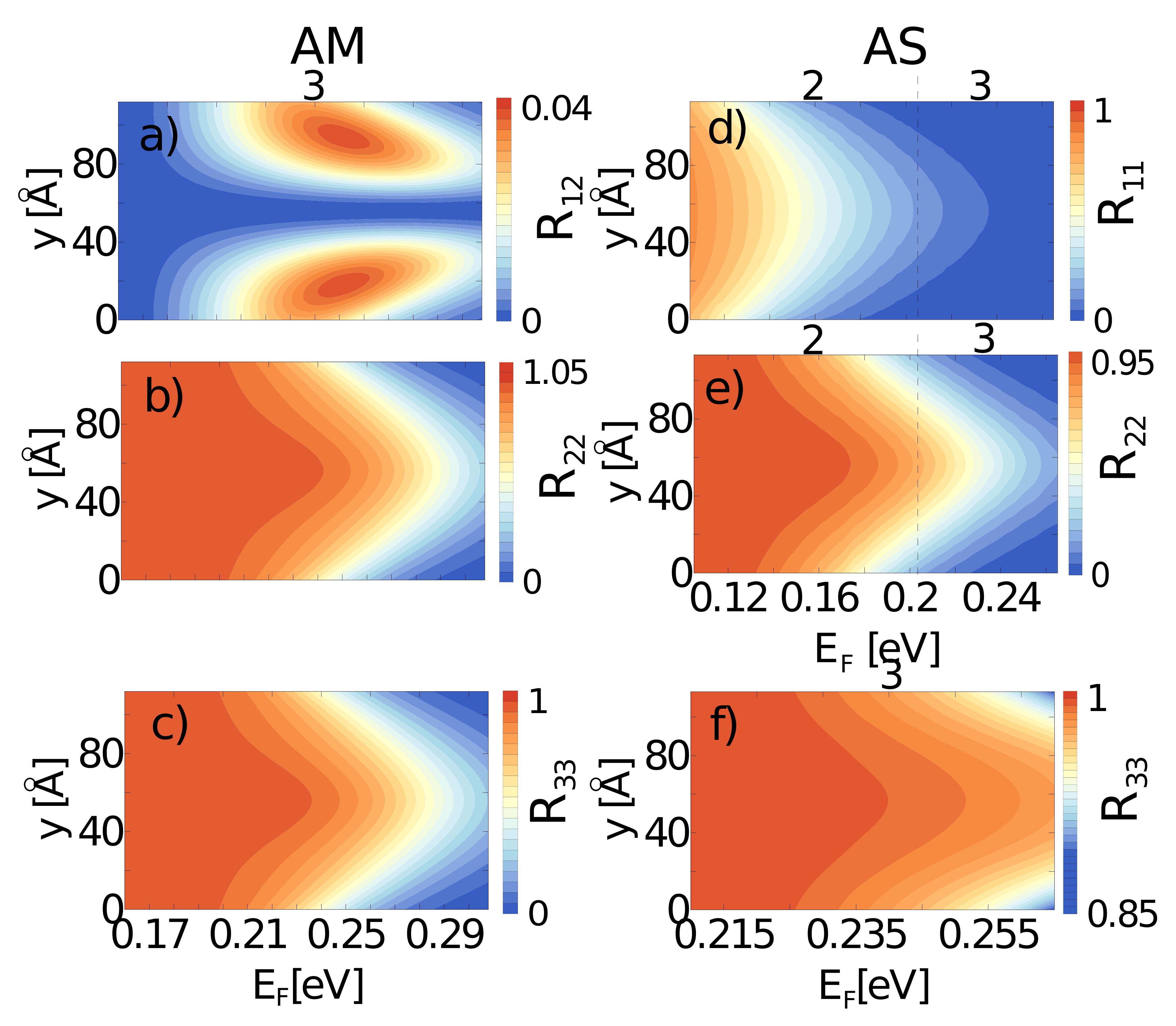}
\caption{Intersubband scattering probabilities for armchair metallic (a,b,c) and semiconducting (d,e,f) nanoribbons for  $V_t=0.2$ eV, $d=40a$. The number of subbands at the Fermi level is given above the plots. For the metallic ribbon (a-c) $R_{11}$ is zero. } \label{armchair}
\end{figure}

The armchair edges of the ribbons strongly mix the valleys and the zero point energy appears only for a single wave vector in the dispersion relation for metallic
channels [Fig. \ref{disprel}(d)] or a single extremum per energy band is observed [Fig. \ref{disprel}(g)] for the semiconducting channel.

For the semiconducting armchair ribbon we find that the tip strictly blocks the current for any low $E_F$  [region '1' in Fig. \ref{armchairskan}(b)] while the metallic ribbon ignores the presence of the tip in the single-subband transport conditions  [region '1' in Fig. \ref{armchairskan}(a)] . In contrast to the case of the zigzag ribbon, the conductance is a smooth function of both $E_F$ and the tip position.
The low-energy results  are given in Fig. \ref{lowestsb} with $T=1$ for the metallic ribbon for both $E_F=0.07$ eV and $E_F=0.13$ eV,
irrelevant of the tip position. The metallic armchair ribbon possesses a perfectly conducting channel due to the pseudovalley symmetry \cite{wurm},
conserved at low energy, hence the absence of backscattering in the lowest subband transport.
For the semiconducting armchair ribbon the tip blocks the current for any tip position above the ribbon [$E_F=0.07$ eV in Fig. \ref{lowestsb}(a)]
or reduces it strongly, particularly near the center of the ribbon [$E_F=0.1$ eV in Fig. \ref{lowestsb}(b)].
In general, the most effective backscattering channel is the one within the same subband [$k_{2+}\rightarrow k_{2-}$, $k_{3+}\rightarrow k_{3-}$] and for the tip above the center of the ribbon.

At higher energies, for the semiconducting ribbon, only the backscattering within the same subband is non-negligible  [Fig. \ref{armchair}(d-f)], including the non-zero channel in the lowest subband.
For the metallic ribbon the backscattering within the lowest subband is absent due to the pseudovalley symmetry \cite{wurm}. Scattering in the same subband [Fig. \ref{armchair}(b-c)] is the strongest as
in the semiconducting case and it is most effective for the tip above the center of the ribbon [Fig. \ref{armchair}(b-f)].
For the metallic ribbon a non-zero scattering between the subbands is also found [Fig. \ref{armchair}(a)] but only for the tip near the edges of the ribbons.
In Fig. \ref{lowestsb}(c) we plotted a cross section of Fig. \ref{zigzagtip}(c) and Fig. \ref{armchairskan} for Fermi energies in multiple subband transport regime.
For higher Fermi energies, the tip can no longer form the n-p junction at the edge, and the characteristic backscattering of the zigzag ribbon is not observed.

Summarizing, for higher Fermi energies the overall response of the ribbon conductance to the tip position is qualitatively similar for all ribbon types,
i.e. it is smooth and maximal at the center of the ribbon.
The $T(y_t)$ dependence can be approximately put in a $(y_t-D/2)^\alpha$
form with $\alpha$ that strongly varies with the energy for any type of the edge, $D$ being the width of the ribbon.
For higher Fermi energies the response of the ribbon to the tip scanning its surface is qualitatively independent of the type of the boundary, in spite of the fact that the current and density distributions are distinctly different [Fig. \ref{lowestsb}(f,i)].

\begin{figure}[htbp]
\includegraphics[width=0.42\textwidth]{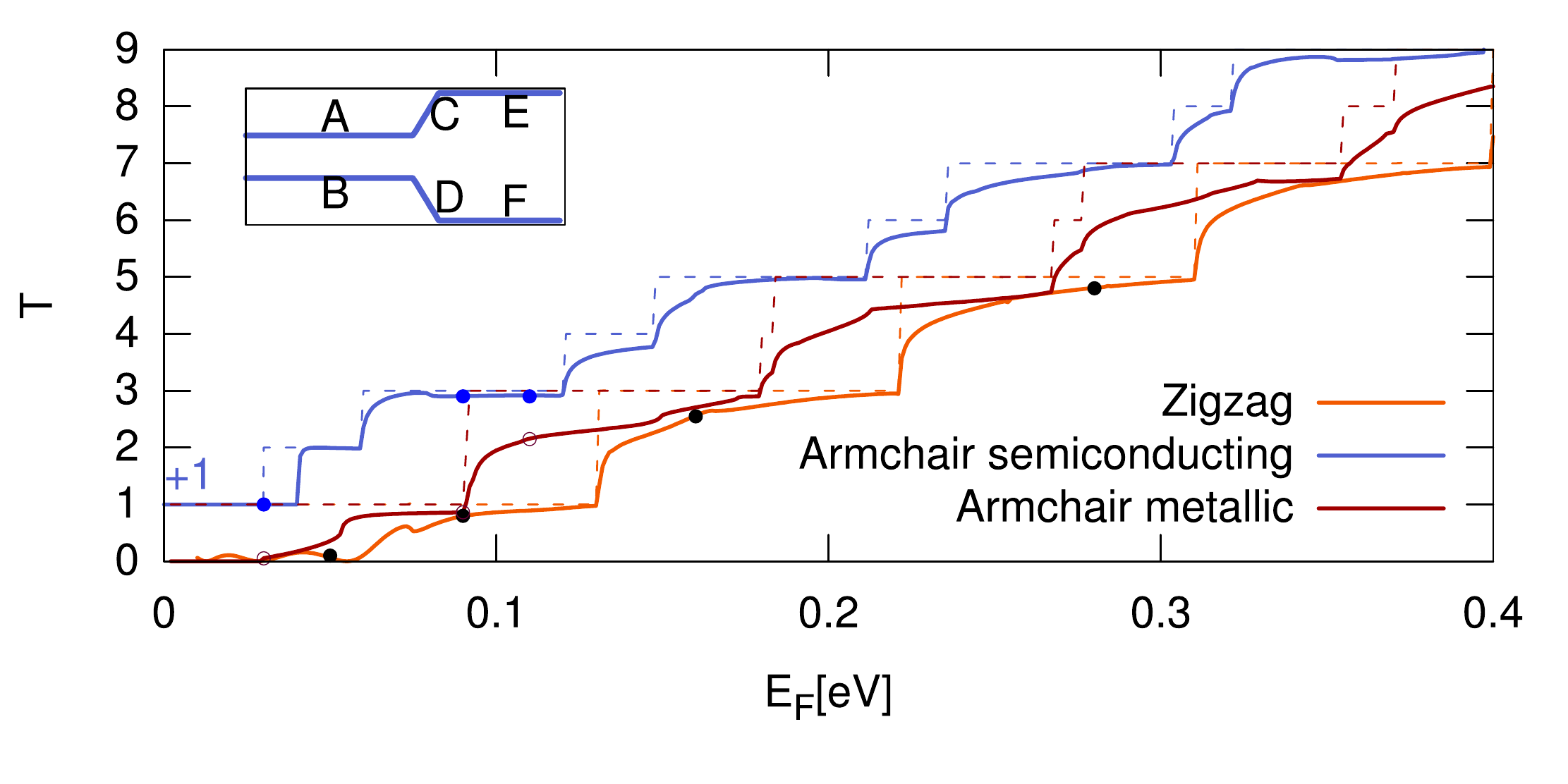}
\caption{Conductance at the contact between channels of varied widths (inset). The channels at the left (A,B) and right (E,F) hand side
of connection and the edge at the variation of the width (C,D) are of the same type : zigzag or armchair.
The lines show the conductance for zigzag, armchair metallic and semiconducting (offset by 1 in $T$) armchair edges of the channels.
The number of atoms across the narrow (wide) channels is 188 (564) for the zigzag system, 161 (363) for metallic armchair channels and 163 (365) for semiconducting armchair edges.
The dots on the lines indicate the points for which the SGM scans of Figs. \ref{zwezigmaps} and \ref{zweziarm} were calculated.
} \label{zwezene}
\end{figure}

\begin{figure}[htbp]
 \includegraphics[width=0.52\textwidth]{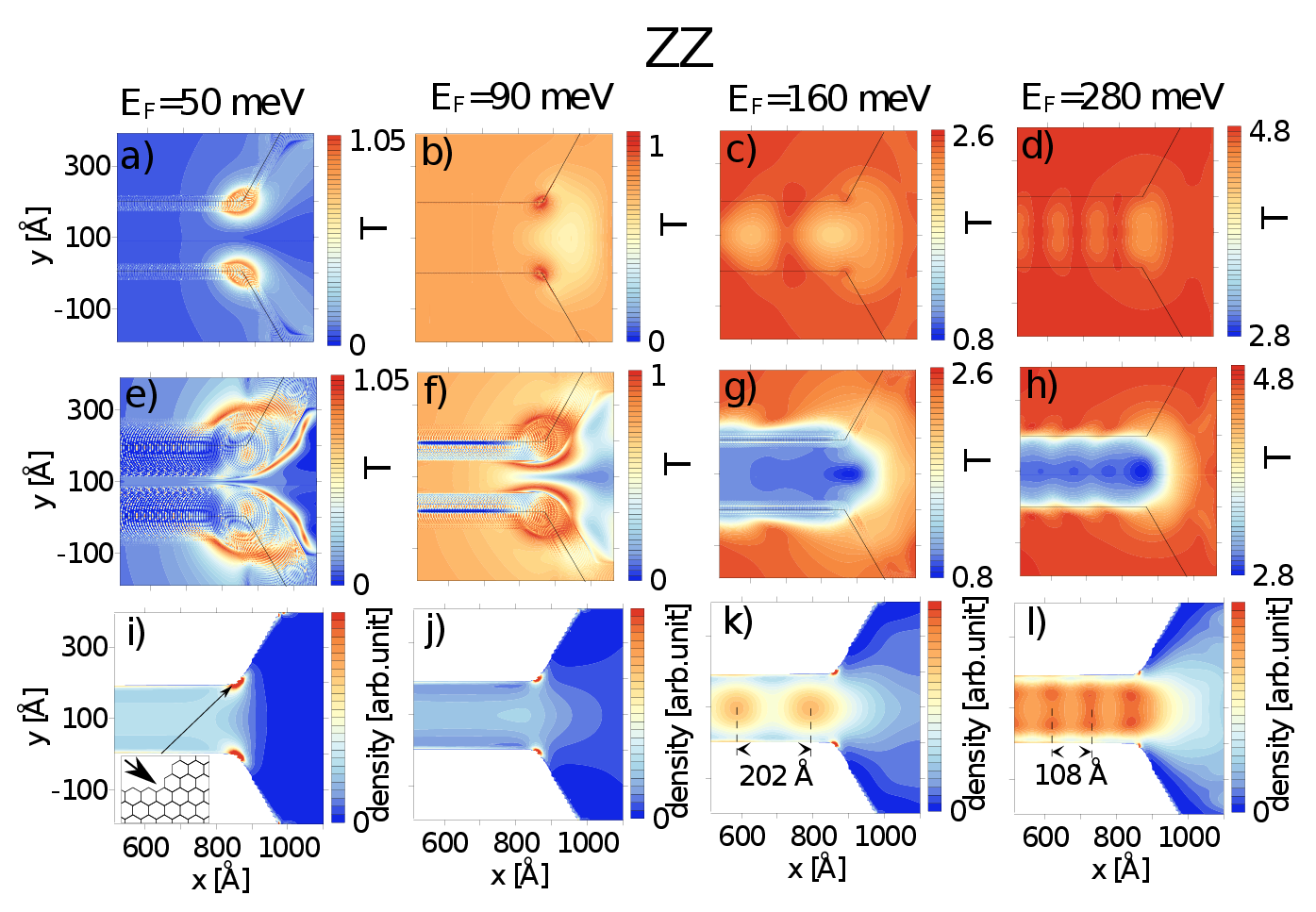}
\caption{Conductance maps for connection of zigzag channels of varied widths for $V_t=0.1$ eV and $d=20a$ (a-d) and $V_t=0.2$ eV and $d=40a$ (e-h).
Panels (i-l) show the scattering probability density for the electron incident from the left in the absence of the tip.
By arrows in (k,l) we indicated the period of the probability density oscillation which corresponds closely to the values
expected for the backscattering within the last subband (see text).
The inset to Fig. 12(i) shows the atomic structure of the turn of the edge with a short armchair segment that induces backscattering at low Fermi energy.
}  \label{zwezigmaps}
\end{figure}

\section{Narrowing of the ribbon}
As an intermediate step towards description of QPCs let us consider the channel of varied width
that is depicted in the inset to Fig. \ref{zwezene}. The channels at both sides of connection ($A,B$ and $E,F$ in Fig. \ref{zwezene}) as well as the linking edges ( $C,D$ in Fig. \ref{zwezene}) are assumed of the same type (zigzag or armchair).
Short segments of varied type of the edge appear at the connections between the horizontal and slanted ends
of the ribbon (see the inset in Fig. \ref{zwezigmaps}(i)).

Figure \ref{zwezene} shows the conductance of the contact as a function of the Fermi energy for both zigzag and armchair edges. The conductance exhibits a very neat quantization
as a function of the Fermi energy. A pronounced quantization of conductance \cite{ccc} is generally difficult to obtain in graphene quantum point contacts (see Introduction).
According to the results of Fig. \ref{zwezene} a channel containing a narrowing is a good candidate for observation of conductance quantization in graphene systems.

\subsection{Zigzag system}
Let us consider the narrowing of the ribbon with zigzag edges.
At low Fermi energy of 50 meV the connection of zigzag edges has a very low conductance of $T=0.006$ [see the orange line in Fig. \ref{zwezene}].
 The ribbons at both sides of connection contain a perfectly conducting
channel, but the inhomogeneity of the width leads to a very strong backscattering.
 At the connection of the zigzag edges a short sequence of an armchair form appears [see the arrow in Fig. \ref{zwezigmaps}(i)], which opens the intervalley backscattering channel.
 The tip potential above any of the two points closes the intervalley backscattering channel, allowing for the electron passage to the wider ribbon [Fig. \ref{zwezigmaps}(a,b,i,j)].
When the tip is close to the zigzag edges, it produces a rapid variation of conductance which is related to formation of the n-p connection at the edge as discussed above. Note, that this applies not only for the narrower ribbon, but also at the widening and at the wider ribbon. Thus the probe resolves the zigzag edges in the entire system.
The strong conductance response near the scattering centers localized at the edges are the counterparts of the conductance halos found in SGM mapping of graphene constrictions \cite{c1}.
For the tip above the center of the narrower ribbon in Fig. \ref{zwezigmaps}(b) the potential at the edge stays below the Fermi level and the presence of the tip is not resolved
in the map which is flat in the interior of the narrow ribbon [Fig. \ref{zwezigmaps}(c,f)].
The insensitiveness of conductance to the tip position in the center of the narrow channel and the rapid variation at the zigzag edge -- are consistent with the properties of the zigzag ribbons for low Fermi energies discussed above.

For higher energies the effects of the edges are no longer present in the conductance maps [Fig. \ref{zwezigmaps}(c,d,g,h)]. Instead we observe -- particularly for the lower potential at the tip [Fig. \ref{zwezigmaps}(c,d)]
an appearance of a periodic oscillation of conductance.
For instance the $T$ maxima in the SGM maps are spaced by $202 \AA$ for $E_F=0.160$ eV
and coincide with the minima of $R_{2'2'}=R_{22}$ backscattering.
The $T$ maxima are spaced by  a characteristic wavelength for the intersubband scattering
 $\Delta x=\frac{2\pi}{|k_{2_+}-k_{2_-}|}$.
The wave vectors are equal to  $k_{2_+}=0.6825  \frac{\pi}{a}$ and $k_{2_-}=-0.6582 \frac{\pi}{a}$ for $E_F=0.160$ eV,
producing spacing $\Delta x$ equal to $202.3 \AA$, in a perfect agreement with the oscillation period found in the SGM map.
For larger $V_t$ and wider $d$ a deviation from this perfect agreement is found due to variation in the wavelengths induced by the external potential of the tip
and  backscattering involving a larger number of subbands. In particular in Fig. \ref{zwezigmaps}(g) for $V_t=0.2$ eV the transport is reduced
to the lowest subband ($T=1$) for the tip above the channel and the oscillations
disappear in the SGM map which takes the form similar to the one found for lower Fermi energy --  Fig. \ref{zwezigmaps}(a,b).

The oscillations of the scattering density observed above for the uniform ribbons [see Fig. \ref{prondy}] resulted from the presence
of the tip but were not resolved by the conductance mapping. For the non-uniform ribbons the oscillations appear
in the scattering wave functions due to the inhomogeneity of the channels already in the absence of the probe [Fig. \ref{zwezigmaps}(k,l)]
and according to the present results they can be resolved by the scanning gate microscopy.

\begin{figure}[htbp]
 \includegraphics[width=0.42\textwidth]{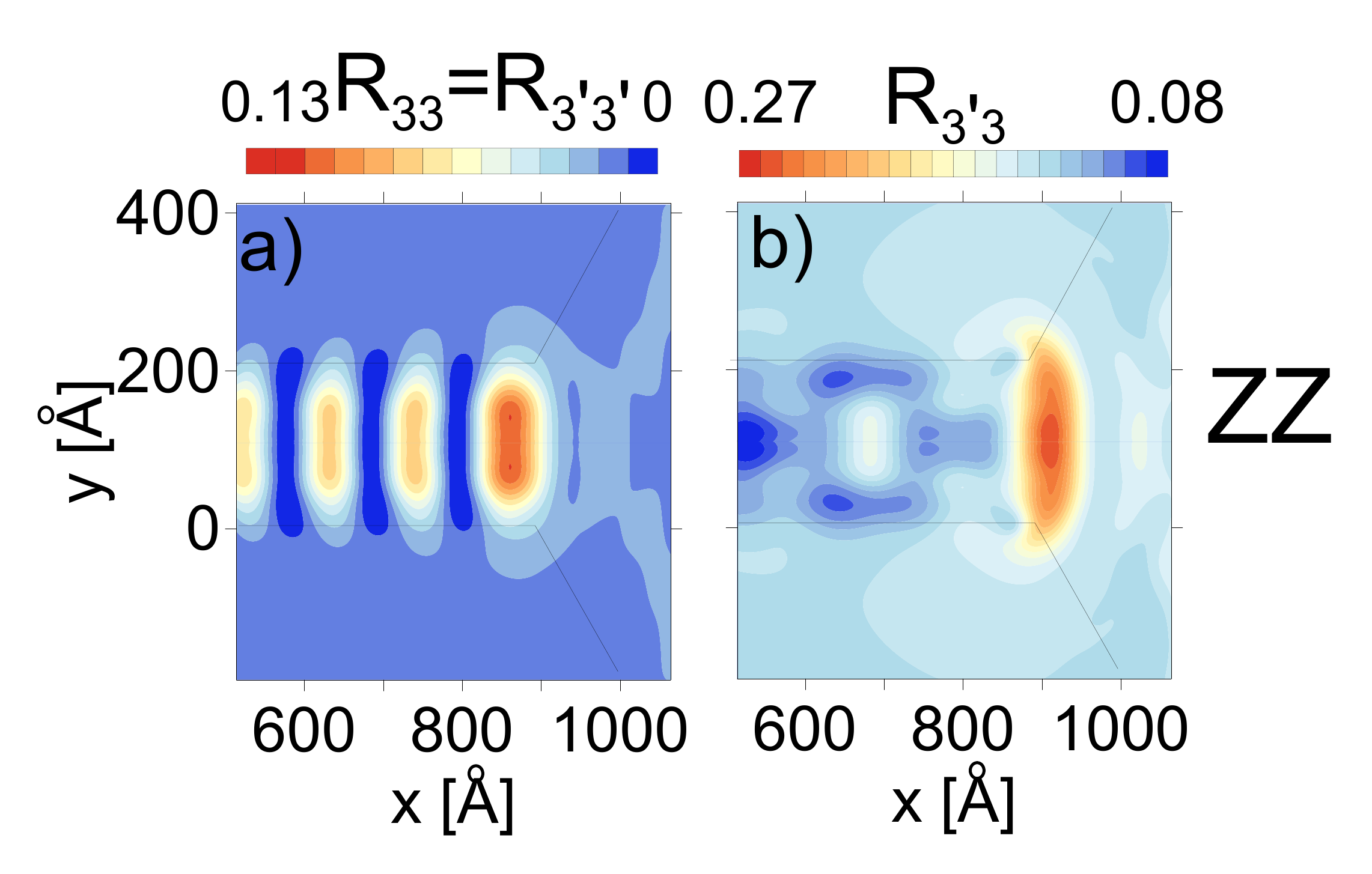}
\caption{Intrasubband (a) and intervalley (b) backscattering for the zigzag narrowing with the Fermi energy of 280 meV  for  $V_t=0.1$ eV, $d=20a$.}  \label{zwezigbac}
\end{figure}

The number of subbands that appear at the Fermi level increases with the energy and several backscattering channels appear.
For $E_F=280$ meV [Fig. \ref{zwezigmaps}(d)] the spacing between the extrema of $T$ map are $\simeq 108\AA$.
The backscattering probabilities are displayed in Fig. \ref{zwezigbac}.
The SGM map $T$ for $E=280$ meV inside the channel [see Fig. \ref{zwezigmaps}(d)] is very well correlated to spatial variation of backscattering $R_{3'3'}=R_{33}$.
The oscillation period resulting from the difference of the wave vectors is $107.23 \AA$ with
a perfect agreement with the spacing of $T$ extrema in the SGM map.
The intervalley scattering [$R_{3'3}$ in Fig. \ref{zwezigbac}(b)] -- gives only an enhanced backscattering at the exit from the thinner channel with no pronounced oscillation within.
We find as a general rule, that -- at least for a moderate tip potential [$V_t=0.1$ eV, $d=20a$] -- the oscillations
of conductance are observed as a function of the tip position inside the zigzag thinner channel and have periodicity corresponding to the backscattering within the highest subband.

\subsection{Armchair systems}

\begin{figure}[htbp]
 \includegraphics[width=0.42\textwidth]{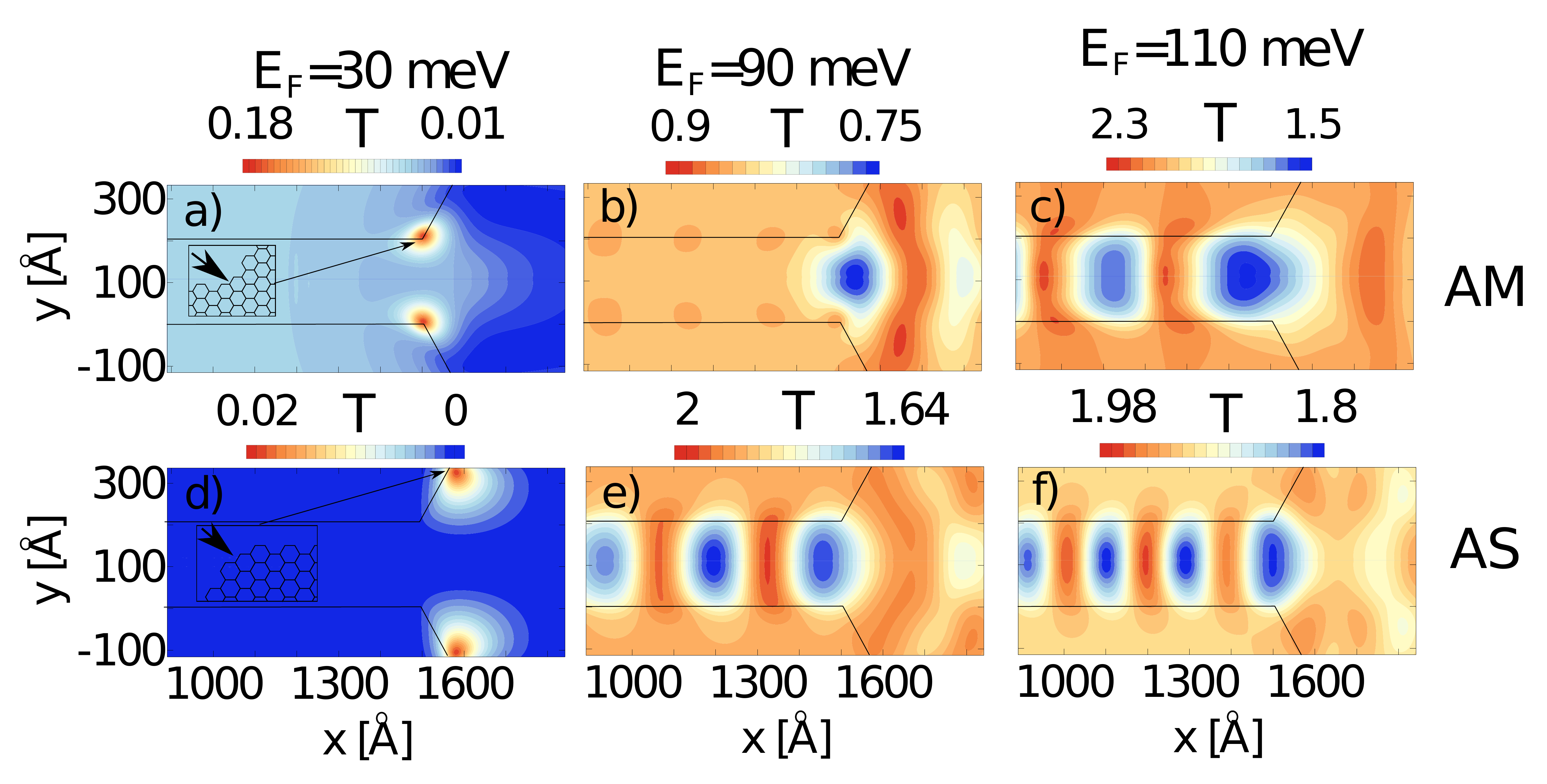}
\caption{$T$ maps for a narrowing with armchair edges: metallic (a-c) and semiconducting (d-f) for $V_t=0.1$ eV and $d=20a$.
The insets in a) and d) display the fragments of the edges near the exit from the narrower channel and the entrance to the wider channel, respectively
where a short segment of zigzag edge appears, leading to the strong backscattering which is neutralized by the probe potential as in Fig. \ref{zwezigmaps} above.}  \label{zweziarm}
\end{figure}

Figure \ref{zweziarm} displays the $T$ maps for the armchair edges of the narrowing [see the inset to Fig. \ref{zwezene}].
At low energy the conductance of both metallic and semiconducting ribbons is nearly zero [Fig. \ref{zwezene}]. Figure \ref{zweziarm}(a,d) indicates
that similarly as for the zigzag system at low energy [Fig. \ref{zwezigmaps}(a,b)] the conductance is raised when the tip is located at the turns of the edges: at connection of the slanted edge to the thinner (for the metallic ribbon Fig. \ref{zweziarm}(a))
or wider (for semiconducting ribbons Fig. \ref{zweziarm}(d)) armchair channel.

For $E_F=90$ meV and the thin semiconducting ribbon we have
already two subbands at the Fermi level [Fig. \ref{zwezene}] and the $T$ map contains a clear oscillation [Fig. \ref{zweziarm}(e)] of periodicity characteristic to the backscattering in the uppermost subband.
For the metallic [Fig. \ref{zweziarm}(b)] ribbon at this energy the thinner ribbon is still in the single subband transport conditions,
no oscillations occur -- exactly as for the zigzag ribbon for low Fermi energy [see Fig. \ref{zwezigmaps}(a-b,e-f)].
For armchair systems  at $E_F=90$ meV no edge effects are observed and the backscattering occurs only for the tip near the exit from the thin channel.
The oscillations for the metallic ribbon occur when the next subbands enter below the Fermi energy [Fig. \ref{zwezene}] -- see Fig. \ref{zweziarm}(c),
with the periodicity characteristic generally to backscattering within the highest subband  with the period of $\frac{2\pi}{|k_{l+}-k_{l-}|}$ for $l$
being the index of the last subband. There are exceptions to this rule,
in particular when the subband is very close to the Fermi energy the backscattering of the subband can be nearly complete, then the one
which is observed in the oscillation is not the highest but the next lower one.

\section{Quantum Point Contacts}
\subsection{Long narrowing of QPC}

\begin{figure}[htbp]
 \includegraphics[width=0.48\textwidth]{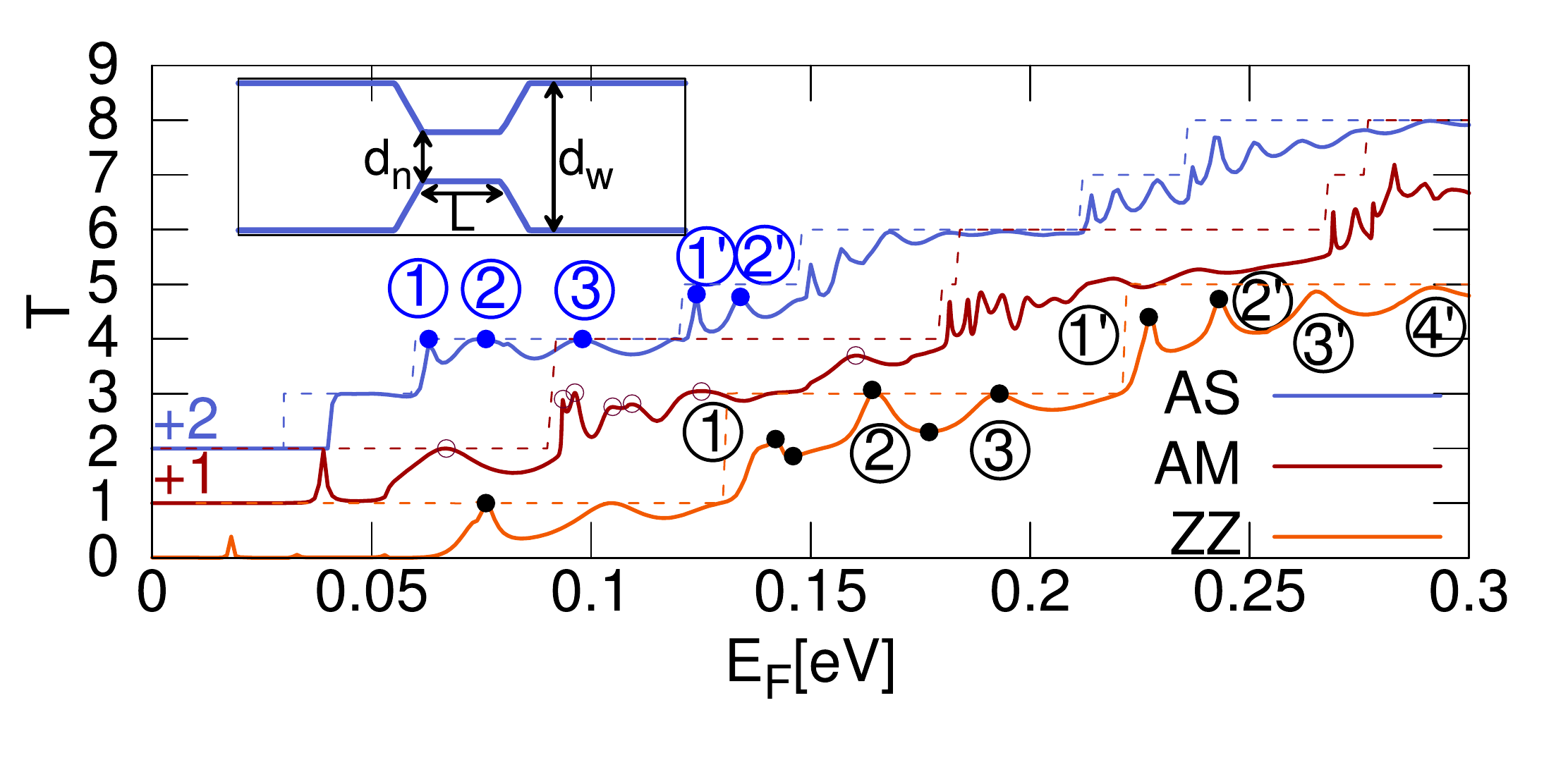}
\caption{Transfer probability as a function of Fermi energy for QPC with long constriction (inset). The dashed lines represent the transfer probability in graphene ribbon of the same width as the constriction. The conductance
for armchair ribbons is shifted by 1 for the metallic system and 2 for semiconducting system. The lengths indicated in the inset are
$L=577$ \AA, $d_n=199.26$ {\AA} and $d_w=447.72$ {\AA} for the armchair semiconducting ribbon, $L=551$ \AA, $d_n=196.8$ {\AA} and $d_w=445.26$ {\AA} for the armchair metallic ribbon and $L=320$ \AA, $d_n=198.8$ {\AA} and $d_w=599.24$ {\AA} for the zigzag ribbon. }  \label{QPCdlugie}
\end{figure}

\begin{figure*}[htbp]
 \includegraphics[width=0.82\textwidth]{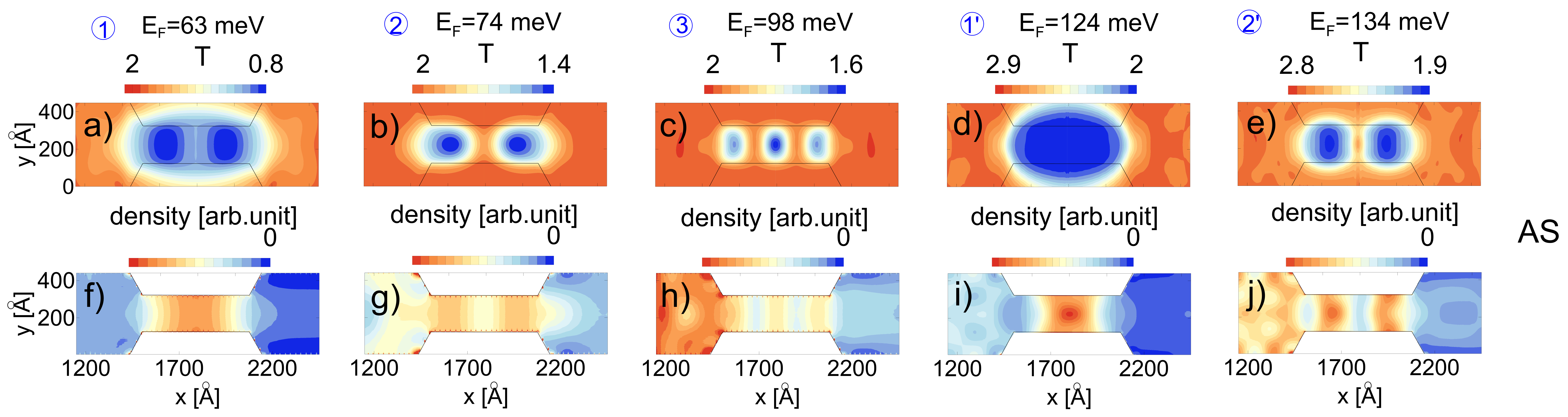}
 \includegraphics[width=0.82\textwidth]{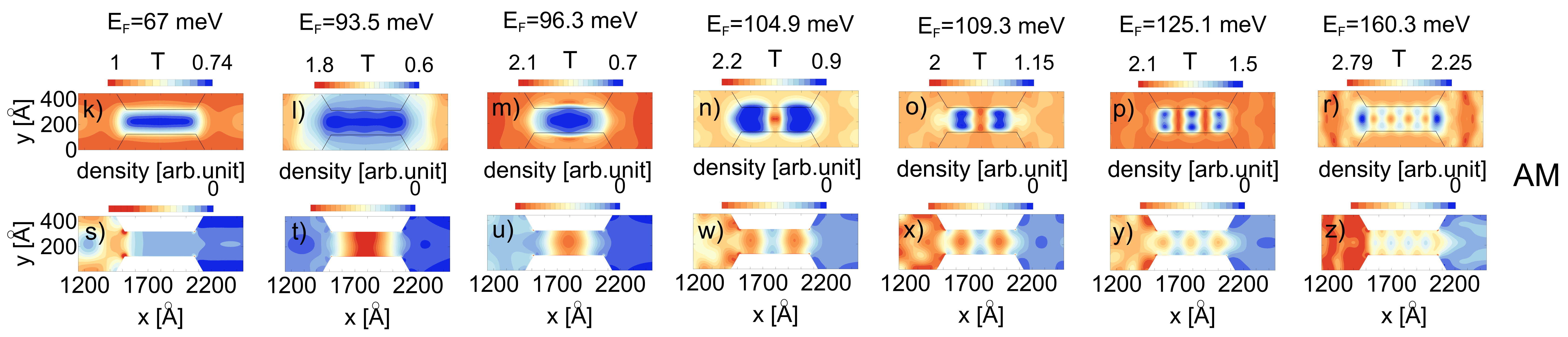}
 \includegraphics[width=0.82\textwidth]{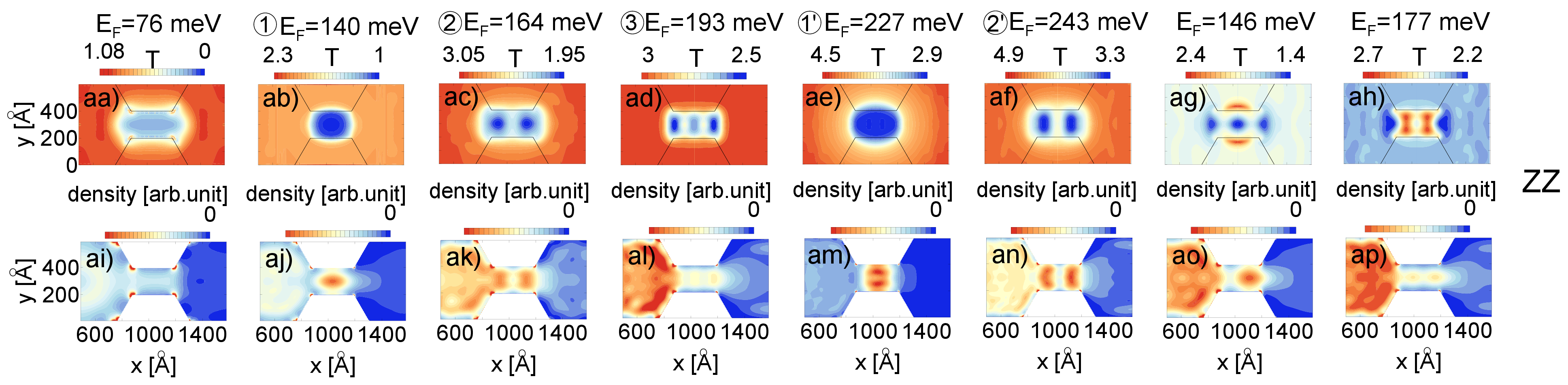}
\caption{$T$ maps for a long QPC with zigzag edges (a-d) and scattering densities (e-h) at the Fermi energies corresponding to the maxima of transmission probability. Parameters of the tip: $V_t=0.1$ eV, $d=20a$.}  \label{dlZig}
\end{figure*}

We are now ready to proceed to description of SGM maps for quantum point contacts formed
within the graphene ribbons. We first consider a system based on two narrowings of the channels as described
above, in the form given in the inset to Fig. \ref{QPCdlugie} with a finite length $L$ of the narrower ribbon. We assume that the narrower ribbon has a width $d_n\simeq$ 20 nm and consider $L$ between 32 nm and 58 nm,
for the transverse size of the wider ribbon between 44 nm and 60 nm.
The conductance of all the considered ribbons -- zigzag and armchair -- both semiconducting and metallic --
drops to zero for low $E_F$. The inhomogeneity of the ribbon leads in this way to
appearance of the transport gap \cite{tg} for any type of the edge of the wider ribbon.

The conductance of the QPC with zigzag edges is displayed in Fig. \ref{QPCdlugie} with the solid orange line,
while the dashed orange line indicates the conductance of the narrow part of the system (the number of subbands
carrying the current to the right at the Fermi level).
The conductance of the QPC is generally lower than the conductance of the narrower part with exceptions
of a series of conductance peaks. In Fig. \ref{dlZig} we plotted the scattering probability density [Fig. \ref{dlZig}(ai-ap)] and
the SGM maps [Fig. \ref{dlZig}(aa-ah)] for the zigzag QPC
and a number of chosen Fermi energies marked in Fig. \ref{QPCdlugie} with points.  In the $T$ maps we find a pronounced variation of conductance maps within the narrower part of the ribbon.
Outside the constriction the conductance maps are almost flat. No conductance variation of visible in the scale of the variations within the narrower channel.
For the low-energy peak $E_F=76$ meV -- within a single subband of the narrow channel -- the conductance map of the system drops
when the probe is located within the constriction [Fig. \ref{dlZig}(aa)].
In the scattering densities and conductance maps we find signatures of backscattering at the turns of the edges as discussed above for the asymmetric narrowing of the channel [Fig. \ref{dlZig}(aa)].
 For higher energies [Fig. \ref{dlZig}(ab-ah)] the corners
correspond to high values of the probability densities but with no counterparts in the SGM map -- again as in the
asymmetric narrowing of Fig. \ref{zwezigmaps}.

For higher Fermi energies, within each segment of $E_F$ corresponding to a fixed number of subbands of the narrow channel (dashed orange line in Fig. \ref{QPCdlugie} we observe
that 1) at the first maximum of $T(E_F)$ at the low energy side the conductance map contains a single minimum along the channel,
2) each subsequent $T$ peak that is higher at the $E_F$ scale corresponds to conductance map with a number of minima increased by 1.
A representative example is given in  Fig. \ref{dlZig}(ab-ad) for the $T$ maps for the first, second and the third $T$ maxima of conductance
[see the numbers near the orange line in Fig. \ref{QPCdlugie}].
 The conductance resonances correspond to an integer number of the scattering wavelengths for the highest subband within the constriction
with maxima observed in the scattering probability density [cf. Fig. \ref{dlZig}(aj-al)]. Once another subband falls below
the Fermi energy (for instance above $E_F=0.2$ eV) another sequence of conductance peaks is formed with the number of maxima
within the constriction starting from 1 again (see the peaks labeled by primed numbers and maps of Fig. \ref{dlZig}(ae,af)].
Beyond the resonances this regular correspondence between the scattering density and the conductance map is lost [cf. Fig. \ref{dlZig}(ag,ah) for energies $E_F=146$ meV and $E_F=177$ meV].

The properties of the QPCs with armchair edges and the width corresponding to a semiconducting dispersion relation turn out to be similar to those of the zigzag system, with pronounced
$T(E_F)$ peaks [cf. blue solid line in Fig. \ref{QPCdlugie}] corresponding to a determined number of maxima
of the probability density correlated with the conductance maps. For the metallic armchair system $T(E_F)$  dependence is more complex [red line in \ref{QPCdlugie}].
In particular we find that the  peaks corresponding to a fixed number of extrema of the probability density
and SGM maps
appear in pairs
[cf. Fig. \ref{dlZig}(l,m) for $E_F=93.5$ meV and $E_F=96.3$ meV for a single extremum, and Fig. \ref{dlZig}(n,o) for $E_F=104.9$ meV and $E_F=109.3$ meV for the double extremum, etc.].

The difference in the properties of semiconducting and metallic armchair systems has its source in the details of  the dispersion relation.
When a subsequent subband for the electron transport is opened within the narrow channel, the subbands of the metallic ribbon are nearly degenerate [see Fig. \ref{disprel}(d) for low $k$]
while for the armchair ribbon the subbands are split near zero wave vector [see Fig. \ref{disprel}(g)]. Thus, for the metallic ribbon we have two close highest subbands of similar
Fermi wavelengths, inducing formation of separate but close resonances for both. The neighborhood of the wavelengths
enhances the intersubband scattering leading to a complex form
of $T(E_F)$ at higher energies.

Concluding, for QPCs with a long constriction the conductance peaks  of $T(E_F)$ dependence are related to resonances
of the highest subband, and the resonant density
within the constriction is resolved by the SGM conductance maps. The tip potential enhances the backscattering
within the highest -- nearly parabolic -- subband. The tip-induced backscattering is most effective in the region
where the scattering probability density is maximal. The zigzag and semiconducting narrowing display a distinct
regularity in the subsequent number of peaks and the form of the SGM map. For the metallic armchair ribbon
the $T(E_F)$ dependence is more complex due to the near degeneracy of the highest subbands near their bottom
and pairs of resonances with similar conductance maps are found. For each system,
outside the constriction the conductance map is almost flat.

\subsection{Short constriction}

\begin{figure}[htbp]
 \includegraphics[width=0.4\textwidth]{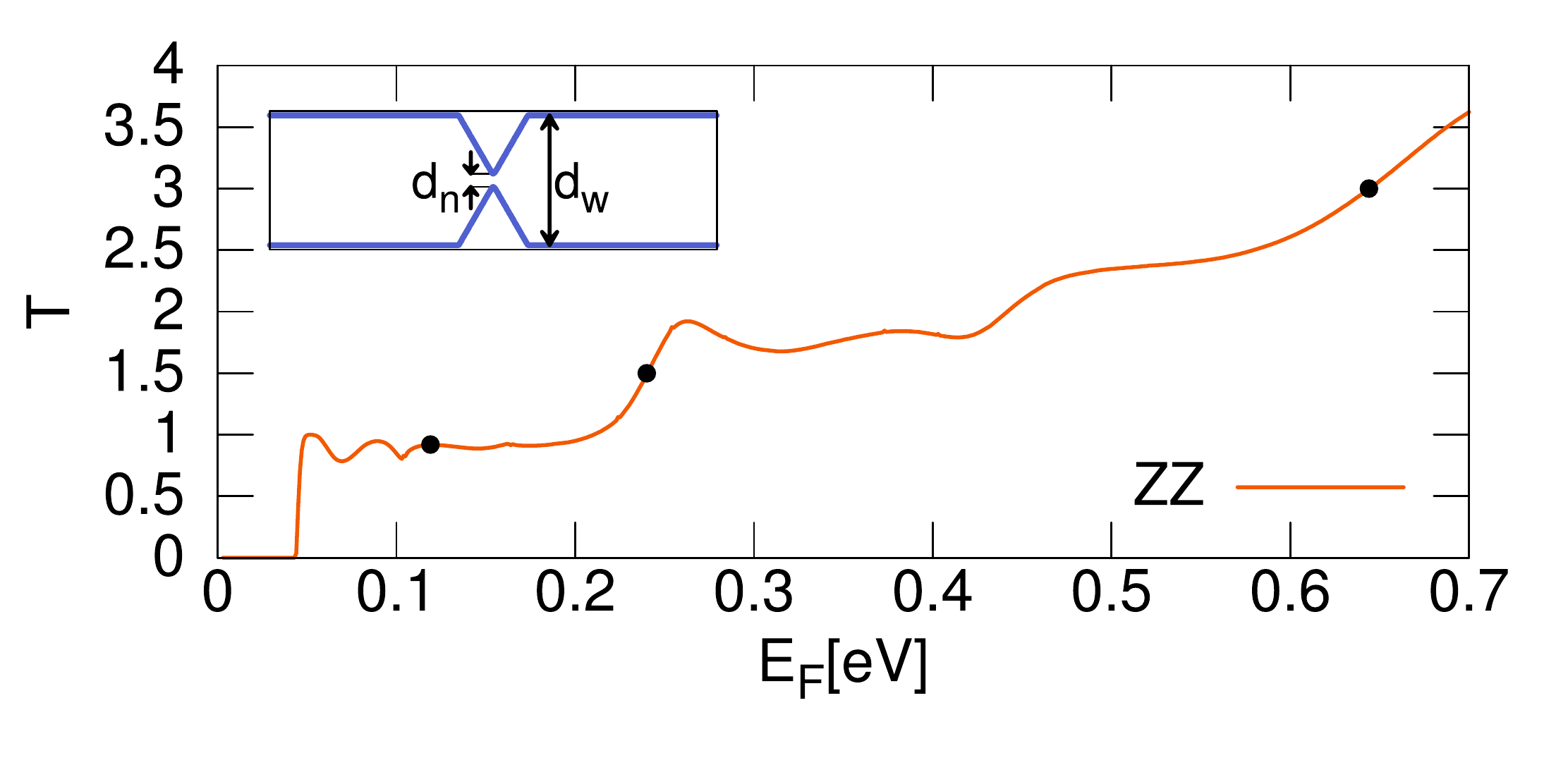}
\caption{Transfer probability as a function of Fermi energy for wedge-shaped QPC (inset). The widths indicated in the inset are $d_n=58$ \AA, $d_w=598.8$ {\AA}.\label{twe}}
\end{figure}

\begin{figure*}[htbp]
 \includegraphics[width=0.62\textwidth]{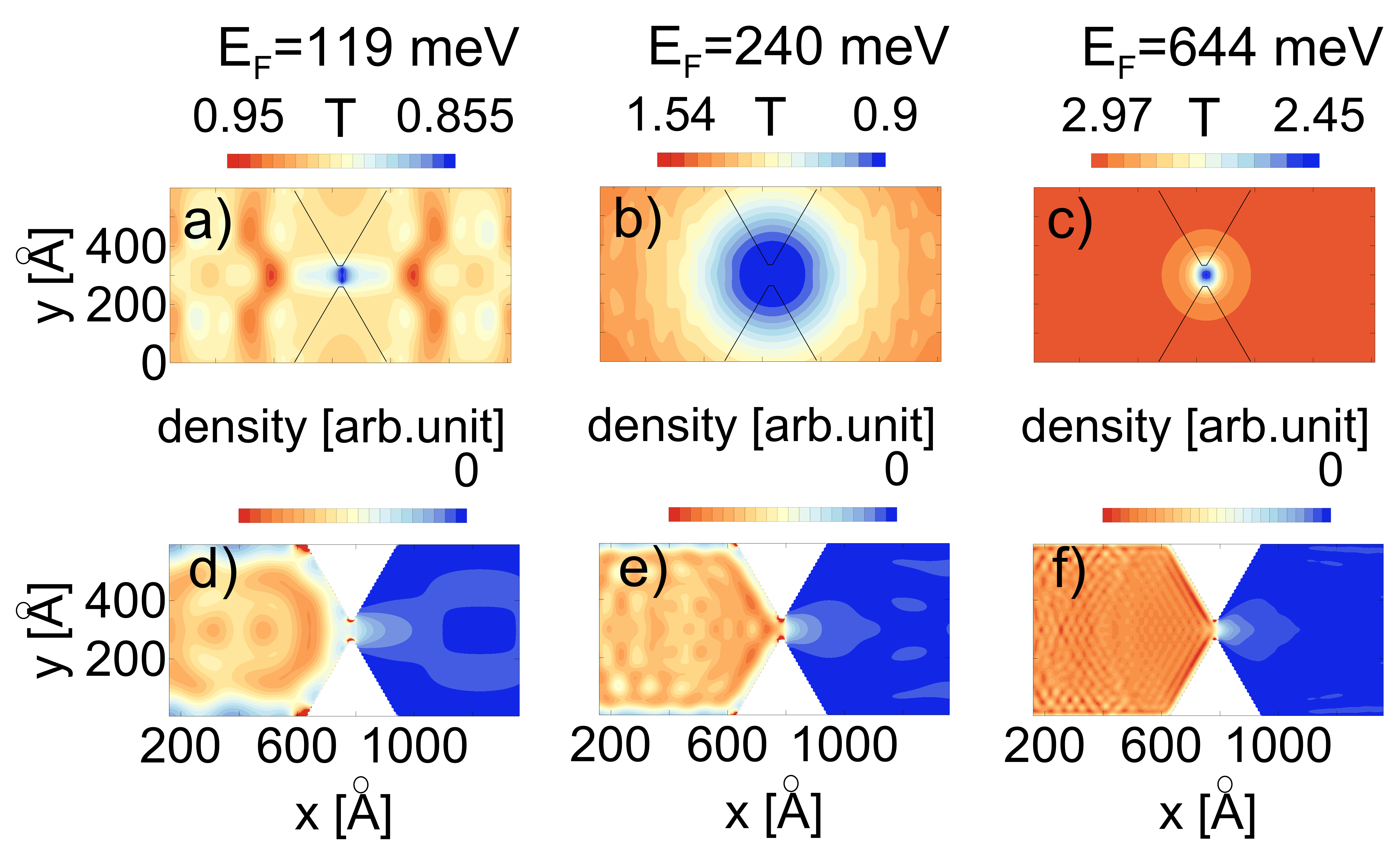}
\caption{T maps for a wedge-shaped QPC with zigzag edges (a-c) and scattering densities (d-f). Parameters of the tip: $V_t=0.1$ eV, $d=20a$.\label{twem}}
\end{figure*}

Let us consider the QPC with a short constriction.
We studied first a wedge shaped constriction of a zigzag ribbon depicted in the inset to Fig. \ref{twe}, with the ribbon narrowing from about 60 nm to about 6 nm.
The conductance of the QPC is displayed in the main panel of Fig. \ref{twe}. The short constriction cannot support any localized resonances in the form discussed above,
and the corresponding series of $T(E_F)$ peaks are not observed.
For any Fermi energy a minimum is found in the SGM map within the constriction and only a low variation of conductance outside the constriction is found
-- see the maps displayed in Fig. \ref{twem}.
Some variation outside the constriction is only observed for low $E_F$ [Fig. \ref{twem}(a)] and it disappears at higher energies in the scale of the variation in the narrow part. [Fig. \ref{twem}(b,c)].

\begin{figure}[htbp]
 \includegraphics[width=0.4\textwidth]{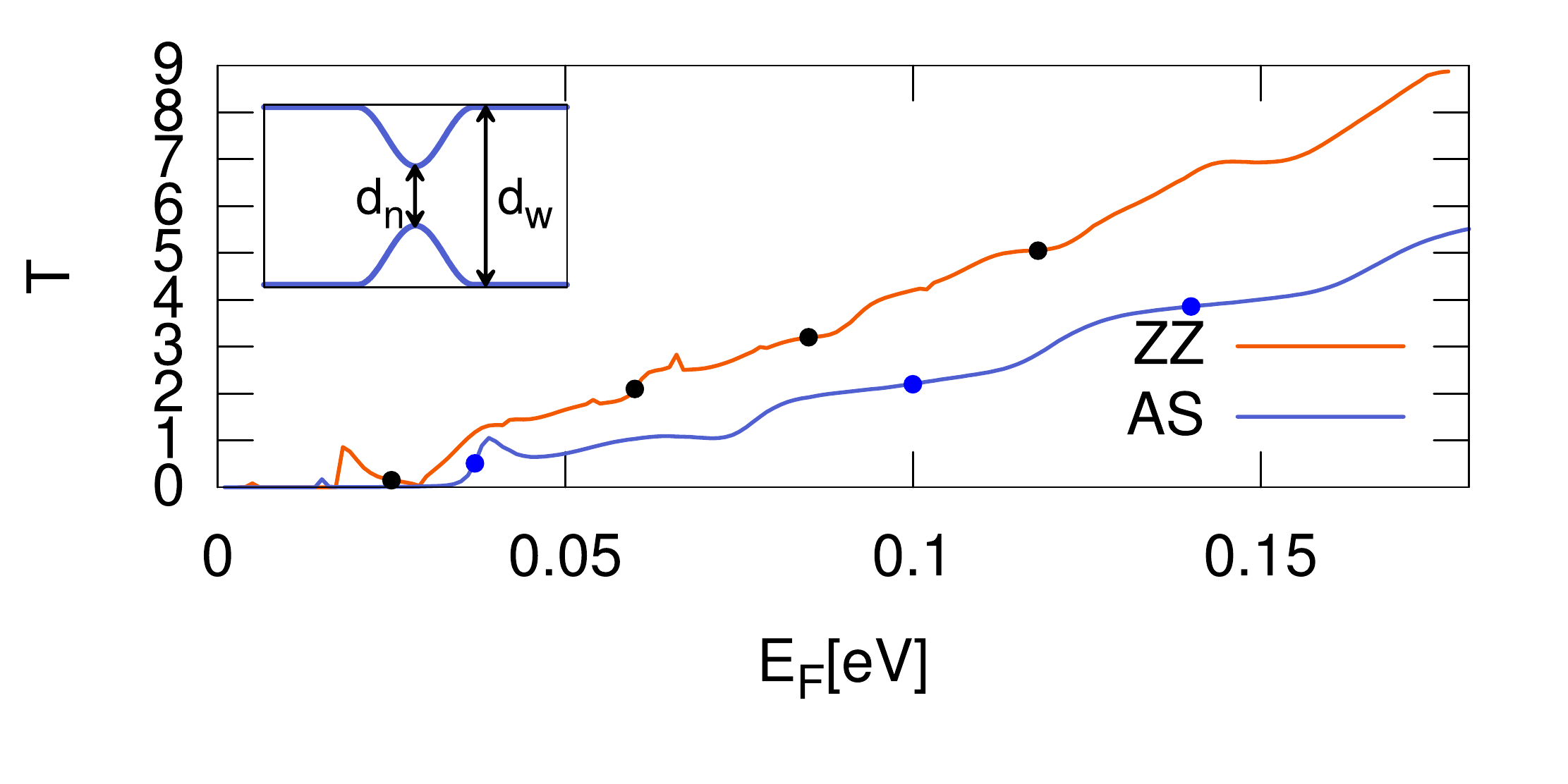}
\caption{Transfer probability as a function of Fermi energy for large QPC (inset). The widths indicated in the inset are $d_n=497$ \AA, $d_w=1498.1$ {\AA} for zigzag ribbons and $d_n=332.1$ \AA, $d_w=998.76$ {\AA} for armchair ribbons.\label{Tcos}}
\end{figure}

The results found for this short, narrow and abrupt QPC are reproduced also for a large smooth at the atomic scale
constriction of a cosine profile \cite{bialorusy}. The system exhibits \cite{bialorusy} a detectable smooth steps of conductance \cite{tombros}
also when the constriction remains quite wide.
The $T(E_F)$ result is presented in Fig. \ref{Tcos},
where we considered wide ($d_w=100$ nm to 150 nm)
 zigzag and semiconducting armchair edges that at the narrowest
part have width of $d_n=33$ nm to 50 nm.

\begin{figure*}[htbp]
 \includegraphics[width=0.82\textwidth]{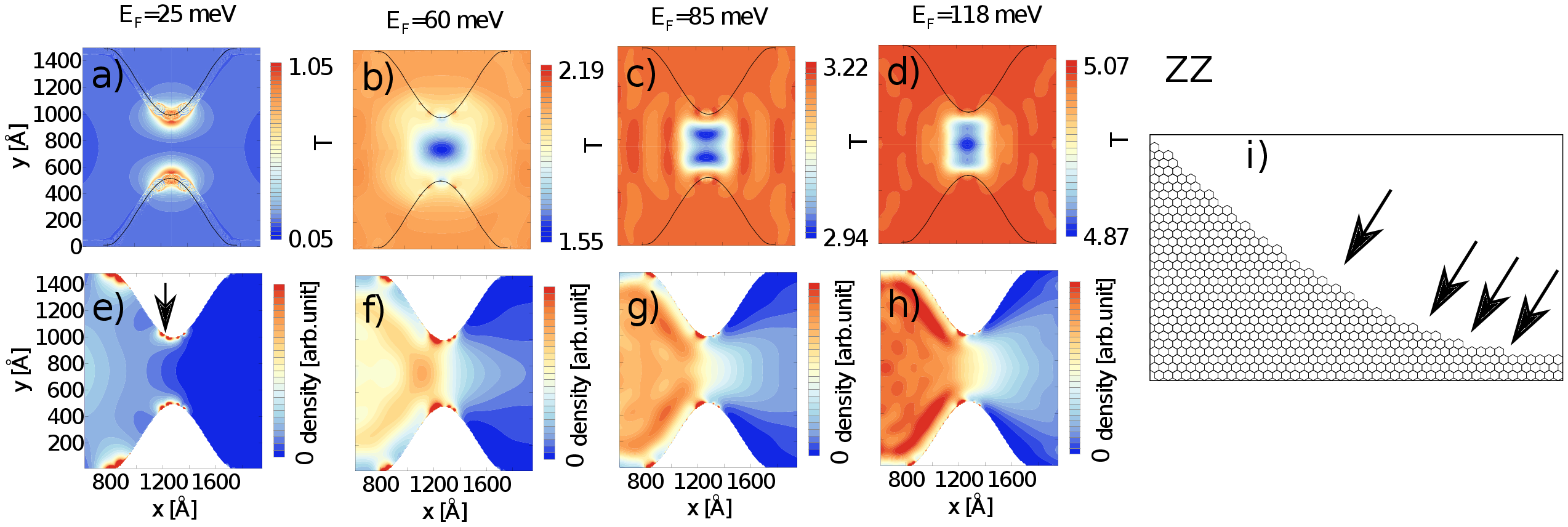}
\caption{$T$ maps (a-d) for a large QPC with mesoscopically smooth edges within a zigzag nanoribbon and the corresponding scattering densities (e-h) in the absence of the probe. Parameters of the tip potential: $V_t=0.1$ eV, $d=20a$. (i) - enlarged fragment of the narrowing indicating armchair segments within the zigzag edge that are mainly responsible for the scattering at low Fermi energy and
that are neutralized by the tip potential [see (a)].}  \label{dlMaps}
\end{figure*}

The conductance maps for the zigzag ribbon are displayed in Fig. \ref{dlMaps}.
In the scattering probability density maps we find maxima along
the constriction [Fig. \ref{dlMaps}(e-h)]. The atomic steps forming constrictions contain multiple
of zigzag-armchair short segments of the edges [Fig. \ref{dlMaps}(i)] -- which as discussed above
-- lead to an effective backscattering. At low energy [$E_F=25$ meV in Fig. \ref{dlMaps}(a)] the tip neutralizes the backscattering when located above these parts
of the edges -- in consistence with the results for asymmetric narrowing
[cf. Fig. \ref{zwezigmaps}(a,b)]. At higher energies [cf. Fig. \ref{dlMaps}(b-d)]
we are left with a minimum of conductance at the center of the constriction and a low variation of conductance in the outside.

\section{Summary and Conclusions}

We have performed an analysis of the conductance response of graphene nanoribbons
and their narrowing to a perturbation introduced by potential of a scanning probe. The study was based on the tight binding implementation of the quantum transmitting boundary method.

We find that at low Fermi energies the response of the zigzag ribbons to the
scanning probe occurs only for the tip above the edges of the channels due to formation of the local n-p junctions allowing for a very effective intervalley backscattering
and producing rapidly varying conductance maps. For the probe above
the interior of the zigzag ribbon the conductance does not react to the perturbation.
The finding is due to existence of the perfectly conducting channel and is also
observed for metallic armchair ribbon irrespective of the position of the probe.
For higher Fermi energies the response of the zizgag and armchair nanoribbons is similar
producing smooth conductance maps as functions of the tip position with the strongest
response in the center of the ribbon.

For the asymmetric narrowing of the ribbon -- which exhibits a very distinct conductance quantization -- the conductance maps at low energy resolve the zigzag edges
and scattering centers near the zigzag-armchair segments which
necessarily appear in atomic steps that form the narrowing. For higher
Fermi energy an oscillation of the scattering probability density is observed,
which in general corresponds to backscattering within the highest subband of the
 narrower channel. According to the present results the oscillation can be resolved
in the maps of conductance response to the probe.

For quantum point contacts similar oscillations of conductance maps are observed for longer constrictions that are connected with resonant peaks of $T(E_F)$ dependence.
Formation of conductance peaks have a regular reentrant character in the energy with subsequent subbands appearing at the Fermi level.
The peaks correspond to scattering probability density and conductance map with a determined number of extrema along the constriction.
Shorter constrictions that do
not support the resonances within the narrowing produce a smooth $T(E_F)$ dependence with SGM maps producing a minima of conductance within the contrictions.
For the asymmetric narrowing -- at the wider part of the ribbon and for QPC outside the constriction no conductance response of the system is observed in the response maps.

\section{Appendix}

\subsection{Channel eigenstates}
The procedure for determining the dispersion relation and the Hamiltonian eigenstates for the nanoribbon channels is described here.
For an infinite, uniform lead the tight-binding Hamiltonian can be put in a matrix\cite{bib3}
\begin{equation}
\mathbf{H} = \left(\begin{array}{c c c c c}
\ddots	&	\cdots		&	0		&	0		&  \\
\vdots	&   \mathbf{H}_{u-1}	&\mathbf{B}^\dagger_{u-1}&	0		& 0\\
0	&   \mathbf{B}_{u-1}	&	\mathbf{H}_u	& \mathbf{B}^\dagger_{u}& 0\\
0	&	0		&	\mathbf{B}_{u}	& \mathbf{H}_{u+1}	& \vdots\\
	&	0		&	0		&	\cdots		& \ddots\\
\end{array}\right),
\end{equation}
where matrix $\mathbf{H}_{u}$ concerns a single elementary cell $u$ [see Fig. \ref{ryszu}] and $\mathbf{B}_{u}$ describes the connection between cells $u$ and $u+1$.
The $\mathbf{H}_{u}$  and $\mathbf{B}_{u}$ matrices have dimensions  $n\times n$ where $n$ is number of atomic orbitals within a single elementary cell.
For ideally periodic nanoribbons forming the channels the matrices $\mathbf{H}_{u}$, $\mathbf{B}_{u}$ are the same for all elementary cells,
and we skip the $u$ index in the following.
The Hamiltonian eigenfunction inside the channel can be written in a form divided on separate cells
\begin{equation}
\boldsymbol{\psi} = \left(\begin{array}{c}
\vdots 		\\
\mathbf{\psi}_{u-1}\\
\mathbf{\psi}_{u}	\\
\mathbf{\psi}_{u+1}\\
\vdots		\\
\end{array}\right),
\end{equation}
where $\mathbf{\psi}_{u}$ is a vector of size $n$.
The wavefunction satisfies,
\begin{equation}
-\mathbf{B} \mathbf{\psi}_{u-1} + (E \mathbf{I} - \mathbf{H} )\mathbf{\psi}_{u} + \mathbf{B}^\dagger \mathbf{\psi}_{u+1}= 0.
\end{equation}
Based on the Bloch form of the wave function [Eq. \ref{blw}] we make the following subsitution:
$$ \mathbf{\psi}_{u-1}=\mathbf{\chi},\quad \mathbf{\psi}_{u}=\lambda \mathbf{\chi}, \quad \mathbf{\psi}_{u+1}=\lambda^2 \mathbf{\chi},  $$
for which
\begin{equation}
-\mathbf{B} \mathbf{\chi} + \lambda(E \mathbf{I} - \mathbf{H} )\mathbf{\chi} + \lambda^2 \mathbf{B}^\dagger \mathbf{\chi}= 0.
\end{equation}
With
$$ \mathbf{\eta}=\lambda \mathbf{\chi} $$
the eigenproblem can be put in form
\begin{equation}
\left[
 \left(\begin{array}{c c}
  0		& \mathbf{I}		\\
  -\mathbf{B}	&E \mathbf{I}-\mathbf{H}	\\
 \end{array}\right)
-\lambda
 \left(\begin{array}{c c}
  \mathbf{I}	& 0			\\
  0		&\mathbf{B}^\dagger	\\
 \end{array}\right)
\right]
\left[
 \begin{array}{c}
   \mathbf{\chi} \\
   \mathbf{\eta} \\
 \end{array}\right] = 0.
\end{equation}
The generalized $(2n\times 2n)$ eigenproblem has $2n$ solutions:
$n$ left-going and $n$ right-going modes -- propagating or decaying.
 \cite{bib3}
For evanescent modes it is straighforward to identify right and left going modes.
For right-going evanescent modes the eigenvalue satisfies $|\lambda_{+,n}|<1$ and for left-going evanescent modes $|\lambda_{-,n}|>1$.
The propagating modes -- for the Bloch waves of Eq. (\ref{blw}) have the form $\lambda_{\pm,n}=\exp(ik\Delta x)$, with a real $k$, hence $|\lambda_{\pm,n}=1|$.
For a given $E$ we look for the values of $|\lambda|=1$ and determine the corresponding wave vectors and the periodic functions $\chi$.
Finally, we evaluate the current flux (\ref{cflux}) and determine the direction of propagation -- the current flux is positive (negative) for right (left) going modes.

\subsection{Determination of the scattering amplitudes}
The scattering amplitudes of Eq. (\ref{eq:left1}) and Eq. (\ref{eq:right}) are determined in the following manner.
We evaluate the coefficients $d_{in}^{l}$ multiplying (\ref{eq:left1}) by a complex conjugate of $\chi^{k_-^{l'} }$:
\begin{equation}
 \langle \chi^{k_-^{l'} } | \psi_{0} \rangle =
\sum\limits_l ( c_{in}^l \langle \chi^{k_-^{l'} } | \chi^{k_+^l}  \rangle 
+ d_{in}^{l} \langle \chi^{k_-^{l'} } | \chi^{k_-^l} \rangle 
  ).\label{eeq}
\end{equation}
Here we denote the inner product in discrete form $\langle A | B \rangle = \sum\limits_j A_j^* B_j $.
Defining vector {\bf A} with ${\bf A}_{l'} = \langle \chi^{k_-^{l'} } | \psi_{0} \rangle $
and matrices  ${\bf B}_{l',l} = \langle \chi^{k_-^{l'} } | \chi^{k_+^l}  \rangle $,
$ {\bf S}_{l',l} = \langle \chi^{k_-^{l'} } | \chi^{k_-^l} \rangle $ Eq. (\ref{eeq})
can be written in a matrix form
\begin{equation}
\mathbf{A} = \mathbf{B} \mathbf{ c}_{in} + \mathbf{S} \mathbf{d}_{in},
\end{equation}
which implies
\begin{eqnarray}
{\bf d}_{in}^l &=&\sum\limits_{l'} ({\bf S}^{-1})_{l,l'} {\bf A}_{l'} - \sum\limits_{l',j}
({\bf S}^{-1})_{l,j} {\bf B}_{j,l'} {\bf c}_{in}^{l'} \nonumber \\
 {\bf d}_{in}^{l'}& =&  \sum\limits_{l'} ({\bf S}^{-1})_{l,l'}  \langle \chi^{k_-^{l'} } | \psi_{0} \rangle -  {\bf c}_{in2}^{l'}.
\label{eq:din}
\end{eqnarray}
From the asymptotic conditions for the wave functions (\ref{eq:left1},\ref{eq:right}) we derive the boundary
conditions for the algebraic set of equations for $c_{out}^l$ and $d_{in}^l$.
Let us consider derivative of the wave function [Eq. (\ref{eq:left1})] at the left end of the computational box \begin{widetext}
\begin{align}
\nonumber \frac{1}{\Delta x} (\psi_{0,v} - \psi_{-1,v} ) =& \sum\limits_l ( c_{in}^l \chi_v^{k_+^l} \frac{1}{\Delta x} (1- e^{-i k_+^l \Delta x} ) + d_{in}^{l} \chi_v^{k_-^l} \frac{1}{\Delta x} (1- e^{-i k_-^l \Delta x} )  ) = \\
\nonumber =& \sum\limits_l ( c_{in}^l \chi_v^{k_+^l} \frac{1}{\Delta x} \Delta_{k_+^l} + d_{in}^{l} \chi_v^{k_-^l} \frac{1}{\Delta x} \Delta_{k_-^l} ) .
\end{align}
Using (\ref{eq:din}) we transform this equation to obtain $ \psi_{-1,v} $:
\begin{equation}
 \psi_{-1,v} = 
\psi_{0,v} - \sum\limits_l \left( {\bf c}_{in}^l \chi_v^{k_+^l} \Delta_{k_+^l} - {\bf c}_{in2}^{l'}\chi_v^{k_-^l} \Delta_{k_-^l} + \sum\limits_{l'} ({\bf S}^{-1})_{l,l'} \langle \chi^{k_-^{l'} }| \psi_{0}\rangle   \chi_v^{k_-^l} \Delta_{k_-^l} \right)
\label{eq:psi_lewy}
\end{equation}

Similarly, for the other end of the computational box we multiply equation (\ref{eq:right}) by the complex conjugate of $(\chi^{ k_+^{l'} } e^{i k_+^{l'} N \Delta x})$:
\begin{equation}
 \langle \chi^{k_+^{l'}} e^{i k_+^{l'} N \Delta x}  | \psi_{N} \rangle =
\sum\limits_l  \langle \chi^{k_+^{l'} } | \chi^{k_+^l}  \rangle e^{i (k_+^l-k_+^{l'}) N \Delta x}{\bf c}_{out}^l
   = \sum\limits_l {\bf S}'_{l',l} {\bf c}_{out}^l,
\end{equation}
hence
\begin{equation}
 c_{out}^l = \sum\limits_{l'} ({\bf S'}^{-1})_{l,l'}  \langle \chi^{k_+^{l'} }  e^{i k_+^{l'} N \Delta x} | \psi_{N} \rangle
\label{eq:cout}
\end{equation}
The derivative at the end of the computational box is
\begin{equation}
\frac{1}{\Delta x}( \psi_{N+1,v} -  \psi_{N,v} ) =  \sum\limits_l c_{out}^l \chi_v^{k_+^l}  e^{i k_+^l N \Delta x} \frac{1}{\Delta x}( e^{i (k_+^l) \Delta x} -1)
\end{equation}
Then, using (\ref{eq:cout}) we obtain:
\begin{equation}
\psi_{N+1,v} =  \psi_{N,v} + \sum\limits_{l,l'} ({\bf S'}^{-1})_{l,l'}  \langle \chi^{k_+^{l'} }  e^{i k_+^{l'} N \Delta x}|  \psi_{N} \rangle
\chi_v^{k_+^l}   e^{i k_+^l N \Delta x} ( e^{i (k_+^l) \Delta x} -1)
\label{eq:psi_prawy}
\end{equation}
The boundary conditions given by (\ref{eq:psi_lewy}) and (\ref{eq:psi_prawy})
are used as expressions for the wave functions outside of the computational box in the
Schr\"odinger equation $H\psi=E\psi$.
Having calculated the wavefunction $\psi$, we can calculate the coefficients $d_{in}^{l}$ from (\ref{eq:din}) and $c_{out}^{l}$ from (\ref{eq:cout}).
The results are next used for evaluation of the scattering probabilities and conductance in the Landauer approach.
\end{widetext}
\end{document}